\documentclass[12pt,preprint]{aastex}
\usepackage{graphicx}

\def\mb{\ifmmode {{\rm B_{435}}}\else
                ${\rm B_{435}}$\fi}
\def\mv{\ifmmode {{\rm V_{606}}}\else
                ${\rm V_{606}}$\fi}
\def\mi{\ifmmode {{\rm i_{775}}}\else
                ${\rm i_{775}}$\fi}
\def\mz{\ifmmode {{\rm z_{850}}}\else
                ${\rm z_{850}}$\fi}

\def\mY{\ifmmode {{\rm Y_{105}}}\else
                ${\rm Y_{105}}$\fi}         
\def\mJ{\ifmmode {{\rm J_{125}}}\else
                ${\rm J_{125}}$\fi}
\def\mH{\ifmmode {{\rm H_{160}}}\else
                ${\rm H_{160}}$\fi}
            
\def\lya{Ly-$\alpha$}
\def\AV{{\it A}$_{\rm V}$}

\def\PMCMC{$\pi$MC$^{2}$}               

\def\numltm0175{53\ } % number of emission line objects with M0<-17.5
\begin{document}
\title{A Link to the Past:\\  Using Markov Chain Monte Carlo Fitting to Constrain Fundamental Parameters of High-Redshift Galaxies}
\author{N. Pirzkal\altaffilmark{1}, B. Rothberg\altaffilmark{1,2}, Kim K. Nilsson\altaffilmark{3}, S. Finkelstein\altaffilmark{4}, Anton Koekemoer\altaffilmark{1}, Sangeeta Malhotra\altaffilmark{5}, James Rhoads\altaffilmark{5}}
\email{npirzkal@stsci.edu}
\altaffiltext{1}{Space Telescope Science Institute, 3700 San Martin Drive, Baltimore, MD21218, USA}
\altaffiltext{2}{George Mason University, Department of Physics \& Astronomy, MS 3F3, 4400 University Drive, Fairfax, VA 22030, USA}
\altaffiltext{3}{ST-ECF, Karl-Schwarzschild-Stra{\ss}e 2, 85748, Garching bei M\"{u}nchen, Germany}
\altaffiltext{4}{George P. and Cynthia Woods Mitchell Institute for Fundamental Physics \& Astronomy, Department of Physics and Astronomy, Texas A\&M University, College Station, TX 77843, USA}
\altaffiltext{5}{School of Earth \& Space Exploration, Arizona State University, P.O. Box 871404, Tempe, AZ 85287-1404, USA}

\begin{abstract}
We have a developed a new method for fitting spectral energy distributions
(SEDs) to identify and constrain the physical properties of high-redshift (${\rm 4 < z < 8}$)
galaxies. Our approach uses an implementation of  Bayesian based Markov Chain Monte Carlo that we have 
dubbed ``\PMCMC.''  
It allows us to compare observations to arbitrarily complex models  and to compute 95\%  credible intervals that  provide robust
constraints for the model parameters. The work is presented in two
sections.  In the first, we test \PMCMC\ using simulated SEDs to not only
confirm the recovery of the known inputs but to assess the limitations of the method and 
identify potential hazards of SED fitting when applied specifically to high redshift (z $>$ 4) galaxies.  
In the second part of the paper we apply \PMCMC\ to 33 $4<z<8$ objects,
including the spectroscopically confirmed Grism ACS Program for Extragalactic Science (GRAPES) Ly alpha sample (${\rm 4 < z <6}$), supplemented by newly obtained HST/WFC3 
near-IR observations, and several recently reported broad band selected  $z>6$ galaxies.
Using  \PMCMC, we are able  to constrain the stellar mass of these objects and in some cases their stellar age and 
find no evidence that any of these sources formed at a redshift  larger than ${\rm z=8}$, a time when the Universe was ${\rm \approx 0.6 Gyr}$  old. 
\end{abstract}

\section{Introduction}
\label{intro}
\indent Without a doubt, high-redshift galaxies are critical to developing a comprehensive picture
of galaxy evolution.  It is no longer sufficient to simply detect and catalog their existence. The next step
is to determine fundamental parameters such as mass, metallicity, extinction and the ages of their
stellar population(s).  Such archaeological reconstructions
are by no means a simple feat and present significant technical challenges not just for the observations
but for the methods used to deduce their properties.  In the local universe, it is relatively straightforward
to determine mass from kinematic observations (i.e. velocity dispersions, rotational velocity) or information
about the stellar populations from stellar absorption lines and nebular emission lines.  However, at high-z,
such observations are impractical, if not impossible with current technology.  This forces us to rely on
broadband photometry, or deep low-resolution spectroscopy of bright emission lines (if we are particularly fortunate)
to derive fundamental parameters.  One popular technique to glean such information is fitting templates of 
stellar populations, computed from either theoretical isochrones or empirical observations, to the spectral 
energy distribution (SED) of distant galaxies.  The galaxy SEDs are simply broad-band photometric measurements 
obtained in as many filters as possible.  Ideally, SED fitting should allow one to estimate the total 
stellar mass of the galaxy, the age and metallicity of the stellar population, and the amount of extinction 
in galaxies at known redshifts.   However, in most cases, the redshifts of the observed galaxies are themselves unknown, 
adding it to the list of parameters to be determined. The total stellar mass of the galaxy is derived from determining 
the best possible scale factor to fit the stellar population template to the observations. In the simplest case of SED
fitting, stellar templates of fixed metallicity are used, along with specific extinction laws (e.g. Calzetti, foreground
dust, mixed dust \& stars), making the minimum number of parameters to fit
only four (age, \AV, mass, and z).  In recent years, many have attempted to fit two stellar populations to the SED,
which increases the number of free parameters to seven \citep{pirzkal2007,Nilsson2011}.  Complicating things further, metallicity
is often left as another free parameter.  This gives us 5 free parameters for single population fits, and 9 free
parameters for two-population fits! Moreover, many of these parameters are degenerate, that is, the effect on the SED colors
can be the same for two or more different parameters.   The implication of this is that, unless some assumptions are made about 
the values of at least a few of the parameters (i.e. independent redshift determination), the size of the parameter 
space to examine is daunting. \\
\indent Undeterred from the seemingly large parameter space to explore, the popularity and frequency of SED fitting has 
increased significantly in the last few years (e.g. \citet{Mobasher2005,pirzkal2007,McLure2009,labbe2010}) due to the 
availability of large catalogs of candidate galaxies obtained from deep surveys.   Templates from \citet{bc03}, \citet{ma05}, 
or the newer, but unpublished \citet{bc10} (hereafter BC03, M05, CB11) are the most commonly used for comparisons.   
\citet{bc03} rely heavily on theoretical isochrones, while \citet{ma05} and \citet{bc10} include empirical spectral data 
(primarily for red supergiants and asymptotic branch stars). Where the SED fitting methods differ is in how they span the 
rather extensive parameter space of 
input template parameters and how they derive errors for their results.  One way to reduce the voluminous parameter space   
is to compute models only over a finite grid of input parameters, where each point on the grid is  a distinct template model SED.  
However, this grid must be carefully selected and
deriving error bars for each of the parameter of the best fitting model is difficult and time consuming.   
Furthermore, using a preset grid of model parameters introduces
a problem similar to 
selecting an appropriate number of bins for histograms. In this case it is the input parameter grid that must be pre-selected and the 
actual choice of parameter
values can affect the outcome of the fit. Ideally, a very fine grid should be used for each parameter, but in practice the size and 
span of parameter grid is kept small enough, and sparse enough, 
to make the computational time manageable. There are several
dangers which can result from this method, such as inefficiency, excluding ranges in parameter space that ultimately may prove the most
realistic, and simply selecting the ``best-fit'' model from a coarsely sampled parameter space. In some cases this can produce seemingly 
implausible results
\citep[i.e. a very old object in a young universe as is shown in ][]{labbe2010,richard2011}.\\
\indent A novel way to circumvent these limitations is to use Markov Chain Monte Carlo (MCMC) methods.  As will be described in more 
detail later in the text, the MCMC method allows for an efficient and full exploration of parameter space in a reasonable amount of time
and computing power. When based on maximum likelihood statistics, MCMC also allows for arbitrary complex models to be applied. 
In this paper, we have implemented an SED ``fitting'' method that relies on the 
principle of importance sampling, similar to what is done in numerical integration of complex numerical systems, but
using an MCMC approach. In this sense, this
method is not really a fitting method and should rather be regarded as a method to exclude or include ranges of model parameters.
Unlike nearly all other SED fitting techniques
which are affected by the degeneracy between model parameters, our method derives independent and separate posterior probability distributions 
for each of the model parameters. This allows for easy identification of the marginalization of some parameters, and more importantly, 
allows us to attach error bars to each of the physical characteristics derived for our sources.  Using MCMC, marginalization does not require a-priori 
knowledge of the parameter probability density functions, nor does it require a complicated multi-dimensional integral.
MCMC has been used in the past \citep{Sajina2006,Nilsson2007,Serra2011} 
and is becoming increasingly popular \citep{Nilsson2011,acquaviva2011}. A comprehensive review of MCMC can be found in \citet{trotta2008}. \\
\indent As newer and more sensitive instruments probe longer wavelengths, making the high-z universe more accessible,
the number of potential galaxy sources has increased significantly, particularly in the last decade.  The goal
of this paper is to describe our  MCMC approach to SED fitting based on modern computational techniques and demonstrate the insights
we gain over more classical approaches using simulated galaxies and real observations. 
We start by describing our MCMC based methodology in Section \ref{method}.
We then test the effect of various models, of the sizes of error bars, and photometric noise in Section \ref{calib}.
In Section \ref{science}, we use the MCMC approach to determine the physical properties of a sample of high redshift galaxies: 
1) the high redshift  Lyman-alpha emitters (4 $<$ z $<$ 6) first spectroscopically identified 
as part of the GRAPES projects \citep[Grism ACS Program for Extragalactic Science,][]{pirzkal2004,pirzkal2007}
including several sources re-observed  using the new WFC3 on HST; 2)  galaxy candidates at z $>$ 7-8 from \citet{labbe2010}; 
and 3) high redshift lensed candidates from \citet{bradley2008,zheng2009,richard2011}.   
In Section \ref{conc},  we summarize our key findings and discuss future work to both improve our techniques and future 
observations. \\
\indent All data and calculations in this paper assume {\it H}$_{\circ}$ $=$ 70 km s$^{-1}$ Mpc$^{-1}$ and a cosmology of  
$\Omega$$_{\rm M}$ $=$ 0.3, $\Omega$$_{\rm \lambda}$ $=$ 0.7 (q$_{\circ}$ $=$ -0.55).  All flux measurements
used and provided in tables are in AB magnitudes.

\section{Methodology of the MCMC Approach}\label{method}
\subsection{Description of the Markov Chain Monte Carlo Method}
\indent Markov Chain Monte Carlo is a random sampling method whereby the entire model parameter space is explored. 
It differs from traditional Monte Carlo methods in that it does not attempt to sample a multidimensional region 
uniformly. It instead aims at visiting a point {\bf x} with a probability proportional to some given probability 
distribution function {$\bf \rm \pi({\rm \bf x})$}. The method will spend twice the time sampling a given sub 
region than a region that is half as likely. 
Given a set of observed data, D, all that is required is the ability to derive the likelihood of that data, given
the model parameters {\bf x}.
In return, one directly recovers the normalized probability 
density function of each input model parameter by observing how often the method sampled a volume {\bf dx}. 
The advantage of the MCMC approach is that the distribution of a single component {\bf x} can 
be examined by marginalizing the other components. The posterior probability function (PDF) can be constructed  by simply creating a histogram of the values of
a given parameter in the MCMC chain, including only values taken after the chain was observed to converge.
We were motivated to implement  our  new SED fitting technique because current methods that rely on reporting 
the best fitting model often fail to fully explore the model parameter space. They also fail to fully capture the 
degeneracy between various model parameters \citep[e.g.][]{labbe2010}. Once the number of model parameters exceeds
three or four, it rapidly becomes  inefficient to fit SEDs using parameter grid based fitting methods. 
This is especially true when errors for each of the fitted parameters
are estimated by repeating the grid based fitting following a pure  Monte Carlo approach, (e.g. each parameter
requires several thousand additional iterations just to obtain errors). 
Our main motivation for implementing our own MCMC fitting was to
be able to consider complex input models, including  the effect nebular escape fraction ($f_{esc}$), or 
models consisting of more than one stellar population  \citep[as was attempted in][]{pirzkal2007}.
It seems natural to apply the statistically sound method of MCMC to 
the world of astronomical SED fitting, especially when some insight can be gained from deriving proper estimates of 
model parameter credible intervals. When used properly, MCMC is a very powerful method to determine range in model 
parameter values that are consistent with the data down to a chosen 
confidence level. This confidence level is  commonly taken to be at least 95.45\% (sometimes described as $2\sigma$), 
although there appear to be a trend to only report results with 68\% ($1\sigma$) confidence. It is out opinion that 68\% 
credible regions are not constraining enough. In our experience, the posterior probability density we derive for model 
parameters are non gaussian and hence the 95\% credible regions are not simply twice as wide as the 68\% ones.
While cumbersome,  we have opted to sometimes quote and plot both 95\% and 68\% credible regions in this paper so that 
the reader can appreciate how the two differ and how one might be mislead when  68\% intervals are examined.

\subsection{Applied Technique}\label{tech}
Our implementation of  a Markov Chain Monte Carlo SED analysis, \PMCMC, is written in Python and based on the freely 
available pyMC module \citep{pyMC}. The use of the \PMCMC\ Python package allowed us to implement the more advanced MCMC 
variants and allows \PMCMC\  to run on various platforms such as Linux and OSX with no modifications. \PMCMC\ 
generates models using \citet{bc03}, \citet{bc10} or \citet{ma05} stellar population libraries, using 
one (SSP), two (SSP2) stellar instantaneous populations, or an exponentially decaying SFH model (EXP).
 We also included the effect of continuum and nebular lines emissions, following the recipe described in \citet{Nilsson2007}.
The input templates that are used with \PMCMC\ are defined only at discrete ages and metallicities and these templates are 
interpolated to produce models with arbitrary stellar age and metallicities. We usually used flat priors for all parameters 
which assign equal likelihood to all allowed parameter values. This is a conservative choice of priors that are appropriate 
for most cases where we have no a-priori information about the nature of the object we are analyzing. We have also used 
non-flat priors, such as semi gaussian for the extinction, to check that our results were not being biased by the choice 
of a flat prior, or by the bounding values we chose for these parameters. We used either linear, natural logarithmic or base ten logarithmic distributions for the stellar ages and found that these did not affect the final posterior distributions. The only additional assumptions that were made are that the stellar ages are constrained to be smaller then the age of the Universe at a given redshift z, and that the young stellar component in the SSP2 models is constrained to be younger than the older stellar population. \\
\indent The model parameters that can be varied are: the populations ages; the relative ratio between the old and 
young stellar populations (parameterized as the fraction of the total stellar mass that is in the form of the older stellar population); the metallicities of the young or old stellar populations; the extinction 
(\AV, based on \citet{calzetti2000} law); the half-life $\tau$ value in the case of exponential (EXP) models; 
and the nebular emission escape fraction, $f_{esc}$. A value of one for the escape fraction results in no nebular emission while a value of zero results in the maximum amount of nebular continuum and line emission in the simulated spectra.  
The stellar mass is computed deterministically by scaling each computed model to the observations, similar to what was done in 
\citet{Nilsson2011} and \citet{acquaviva2011}. When computing the mass scale and likelihood between model and observations, we 
take into account photometric bands in which flux is not detected and the upper limit to the flux in these bands.  The upper 
limit to the flux in these bands are used as part of a penalized likelihood similar to what was done in \citet{Lai2007} and 
\citet{pirzkal2007}. We have also allowed for the stellar mass to be varied as a stochastic parameter but saw no effect in doing so, expect that the
chains could sometimes take longer to converge.\\
\indent The inputs to \PMCMC\ is a text file that contains: 1) the measured AB magnitudes or fluxes 
(magnitudes are converted to fluxes internally by \PMCMC); 2) errors (real, or upper limits); 
and 3) the list of all parameters with ranges to probe. About 10 iterations per second can be generated using 
\PMCMC\ with today's typical desktop computer. A 10000 iteration chain can be generated in about 30 minutes, 
assuming a step rejection of about 50\%, or about 2 hours per object when running multiple chains of a few 
tens of thousand iterations each. The stepping method found to work best is the  Metropolis step, as 
described in the pyMC documentation.  We have also had equally good results using an Adaptive Metropolis step once the chain had converged. 
The MCMC approach is highly parallelized and we used \PMCMC\ on both a Linux cluster of 88 processor 
and a 36 node OSX Xgrid cluster.

\subsection{MCMC and convergence}\label{convergence}
\indent The Bayesian foundation of MCMC guarantees that credible regions can be derived if a chain has converged 
and if the MCMC chain(s) contains a sufficient number of iterations. There is however nothing inherent in the method 
that ensures that the chain actually  converges towards the best fit to the observations. It is the responsibility of 
the user to check the quality of the fit and convergence of each MCMC chain. We used several methods to do this, some 
empirical and some based on statistical methods. To empirically check for convergence we  started by simply inspecting 
the traces and histograms of the individual parameters contained in  each MCMC chain. Examples of such plots are available 
in \citet{pyMC}.  One big advantage of the MCMC approach is that the method allows for several MCMC chains to be 
generated independently and in parallel and be combined later, after they are checked for convergence. We could 
then assemble a final MCMC chain by discarding chains that did not converged or resulted in poor fits 
(with relatively low maximum likelihood) and combining the rest of the chains together. We found it necessary, 
especially in cases where the input models where complex, to produce several dozen independent short 
(few thousand iterations each) MCMC chains. The chains converging to the highest likelihood were then 
restarted and extended to several tens of thousands iterations each. In some cases, we simply generated
new independent chains with initial values set to the currently best fitting model so that convergence 
time was minimized. We found that examining the intra chain standard deviation (for a given parameter in the chain) 
and comparing it to the inter chain dispersion (using several independent chains) allowed us to determine where 
each chain had converged. More formally, we also routinely checked for convergence of a particular chain parameter 
by computing the mean and variance of segments of a chain, from the beginning to the end, following the work of 
\citet{geweke1992}. We also made use of another diagnostic \citep{raftery1995} which allowed us to derive and 
estimate the  number of iterations to discard at the beginning of the chain, and the minimum number of iterations 
required to estimate the credible region of a parameter to a given accuracy. More information about these procedures 
are described in \citet{pyMC}. Throughout this paper, we conservatively produced MCMC chains that were significantly 
longer than the number derived from the Raftery-Lewis method to estimate a 95\% credible interval with a 5\% precision. \\
\indent While painstaking, we must stress that this convergence checking is both crucial and necessary if 
one aims to derive useful credible intervals. Failure to do this leads to biased credible intervals. We 
further address the issue of whether we can actually trust the implied meaning of the 95\% credible 
regions in Section \ref{credible}.

\section{Calibration of \PMCMC}\label{calib}
\indent Before applying  \PMCMC\ to real data, it is useful to examine how well the method work by 
applying \PMCMC\ to a series of simple simulated observations.  The tests presented here do not constitute an exhaustive set, but we  selected them amongst the many tests we performed because they illustrate how well  \PMCMC\  performs. In Section \ref{simple}, 
we start by first examining the effects of the {\it size} of photometric error bars, and for now 
ignoring the real effects of photometric noise that displaces the observations away from the fiducial values.
We then examine the effect of  increasing the number of fitted parameters, using more complex models in Section 
\ref{simple2}. Finally, we add real photometric noise to determine the effect of displacing the observed fluxes 
away from their fiducial values, in Section \ref{noise}.  When generating simulated observations, we used BC03 
models as input, and  produced simulated fluxes in the following bandpasses: ACS {\it F435W}, {\it F606W}, 
{\it F775W}, {\it F814W}, {\it F850LP} filters,  the WFC3 {\it F105W}, {\it F110W}, {\it F125W}, {\it F160W} 
filters and the IRAC 3.6 and 4.5 micron channels.  This choice was driven by the filters used in Section 
\ref{science} where we apply \PMCMC\ to real observations.

\subsection{The Simplest Case: No real noise, increasing the size of error bars only}\label{simple}
\begin{figure}
\includegraphics[width=5.0in]{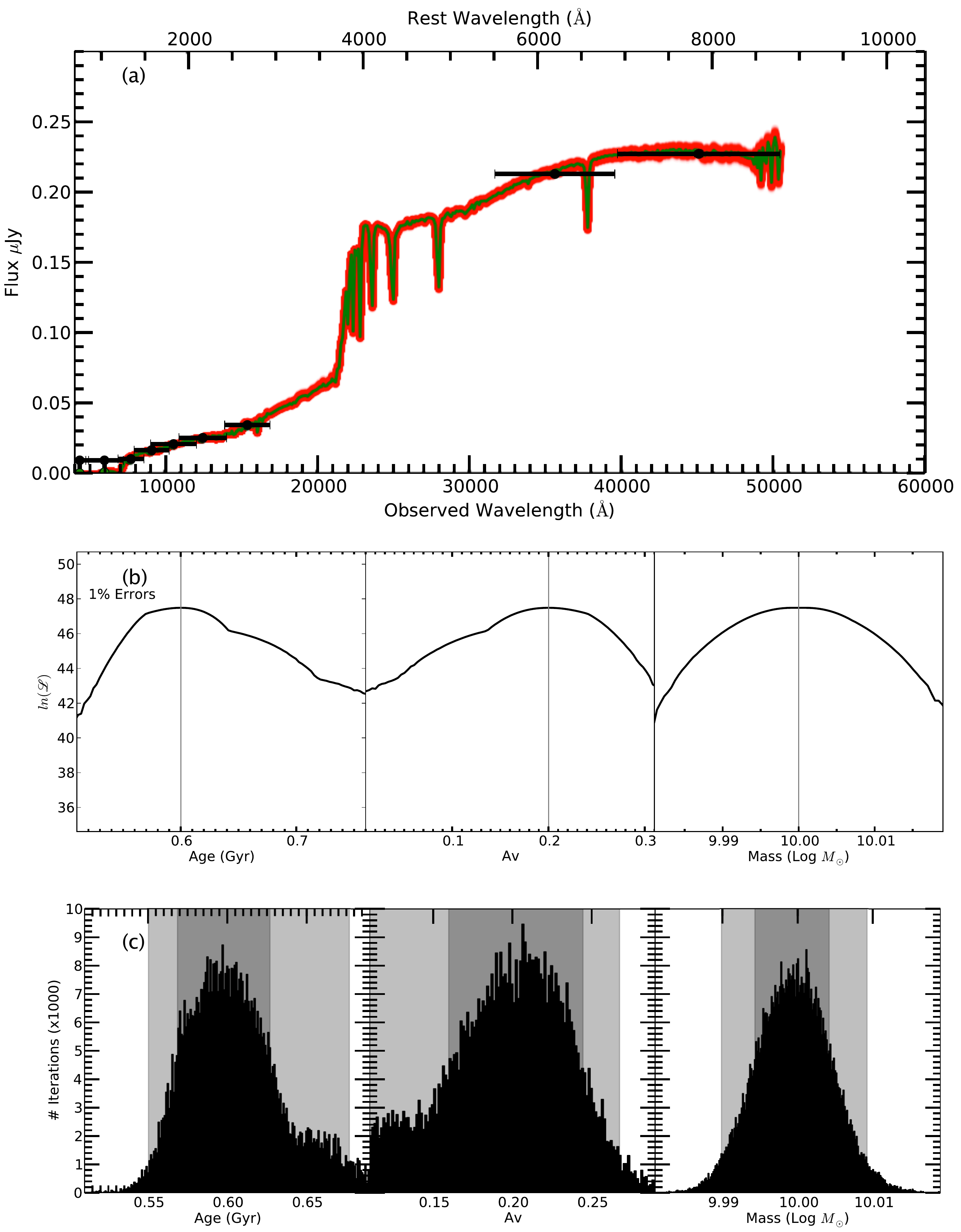}  % fig1.pdf
\caption{Observed simulated single 0.6Gyr old stellar population object at z=4.5 with a mass of ${\rm 10^{10} M_\sun}$ and \AV=0.2, shown with exact photometric values but using 1\% error bars (black error bars) in panel (a). The best fitting model, towards which the chains converged,  is shown using a thick solid green line. Models from the MCMC chain are shown in red and the redder areas correspond to region of the SED where more models happen to lay. Panel (b) shows plots of the log likelihood as a function of individual model parameters. These three plot show how the maximum likelihood (shown with a grey vertical bar) is relatively well defined.
Panel (c) shows  the posteriori probability distributions of the model parameters, as determined from the \PMCMC\ results. We also show the corresponding 66\% and 95\% credible intervals, shown  in light and dark grey respectively.
\label{1percent}}
\end{figure}

\begin{figure}
\includegraphics[width=5.0in]{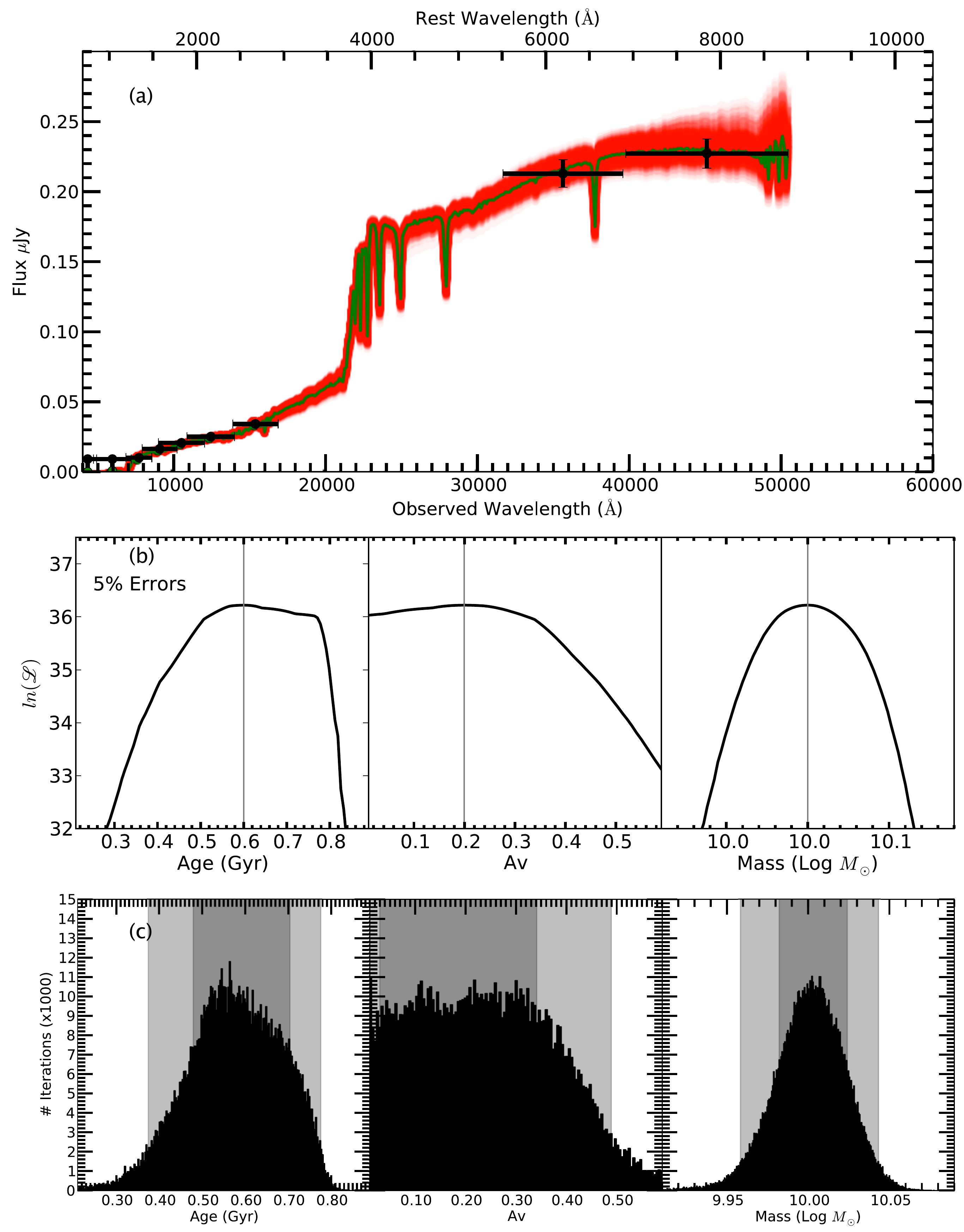} % Fig3.pdf
\caption{This Figure is similar to Figure \ref{1percent} but now shows the effect of allowing for  larger error bars in the input photometry. We now associate a uniform 5\% uncertainty to  each photometric band. The immediate effect of just allowing for larger uncertainties, a broadening of the PDF of each of the model parameters, is clearly shown here.  As larger error bars allow for a larger number of combination of stellar ages, extinction and stellar mass to statistically match the observations. \label{5percent}}
\end{figure}

\begin{figure}
\includegraphics[width=5.0in]{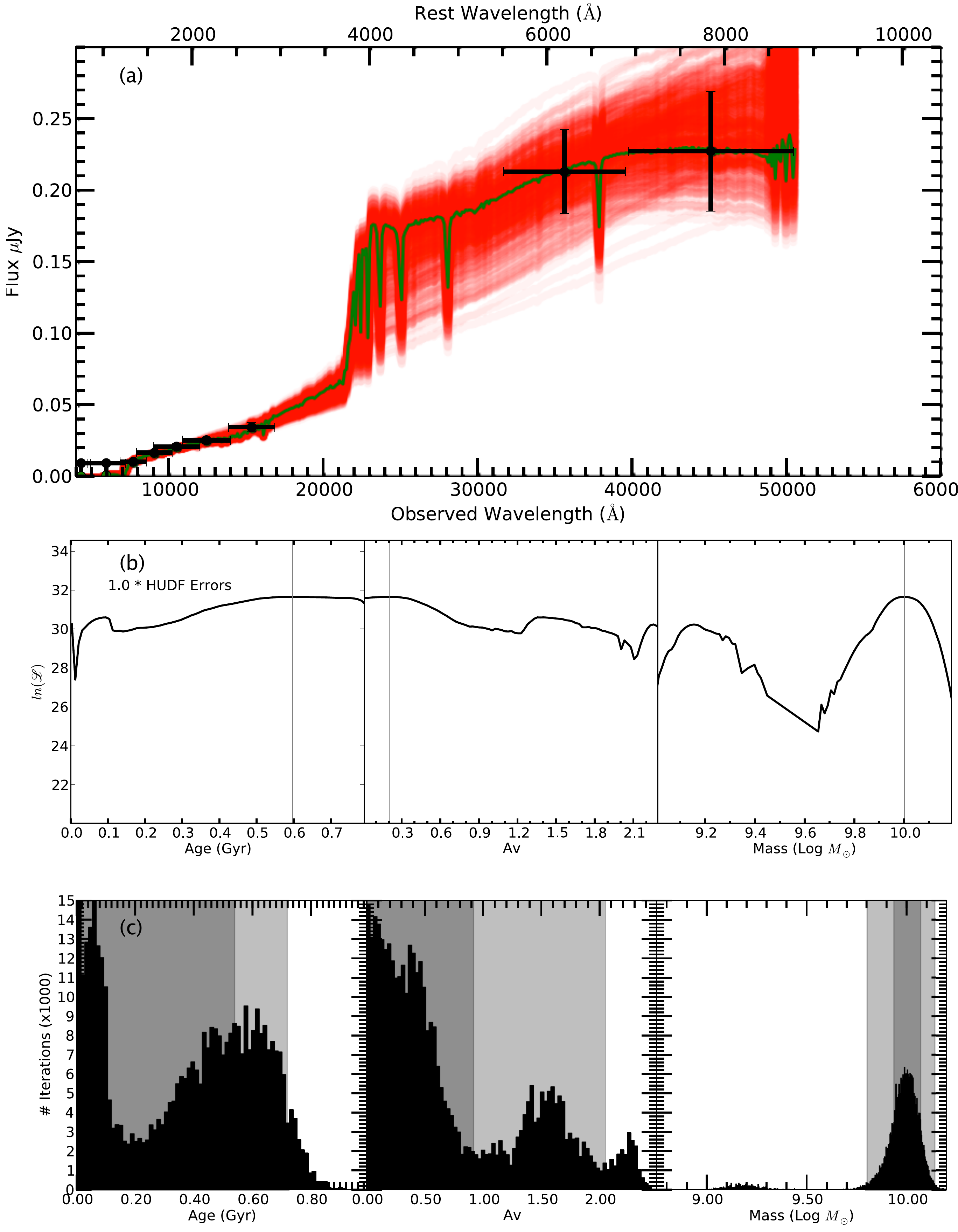} % fig2.pdf
\caption{This Figure is similar to Figures \ref{1percent} and \ref{5percent} but we now adopt more realistic larger error bars in the input photometry. While we assumes a uniform 1\% error in each band in Figure \ref{1percent} and 5\% in Figure \ref{5percent}, we now adopt uncertainties of 5\% in the ACS bands, 10\% in the WFC3 bands and 15 and 20\% in the two IRAC bands. The effect of just assuming larger uncertainties is again very clearly shown here. We now are in a regime where we can see that the PDFs of the stellar ages, extinction, and, to a much smaller extent, stellar mass now take on distinctively non gaussian shapes (panel c). Determining the stellar ages and extinction of this object has now become more difficult as the PDF for these quantities are now very wide. \label{1hudf}}
\end{figure}

\begin{figure}
\includegraphics[width=5.0in]{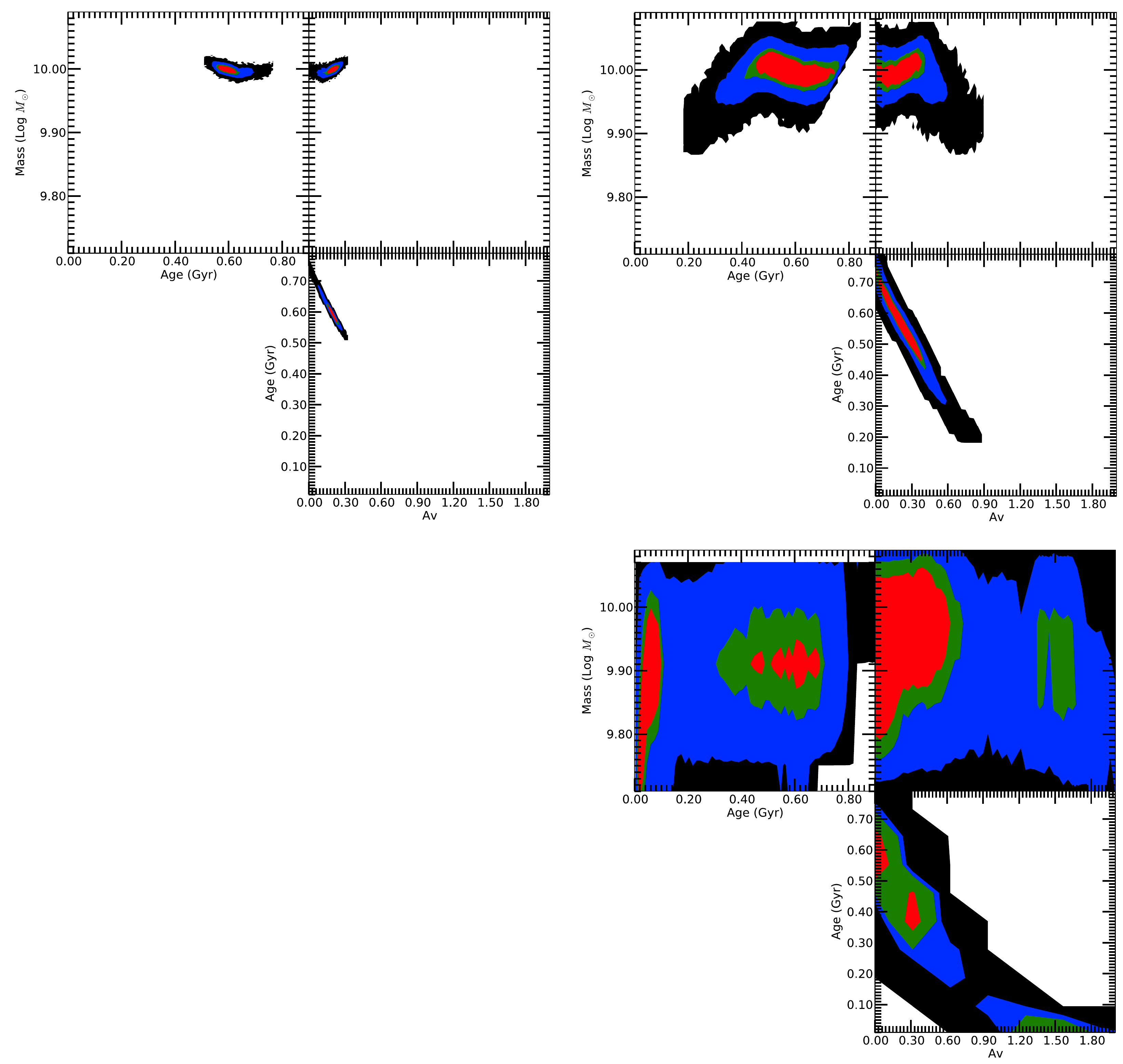} % Fig4.pdf
\caption{The two dimensional model parameters probability distribution obtained with our MCMC SED fitting method and the simulated data shown in Figure \ref{1percent} (1\% uncertainties, top left panels), Figure \ref{5percent} (5\% uncertainties, top right panels), and Figure \ref{1hudf} (Typical HUDF error levels in each photometric band, bottom right panels). The red, green, blue, and black regions are the 50\%, 68\%, 95\% and 99\% credible regions, respectively. All three cases show credible regions that included the fiducial stellar age of 0.6Gyr,  extinction of \AV=0.2 and  stellar mass of $10^{10}  M_\sun$. However the  the size of the credible regions increased and degeneracy between stellar ages and extinction also increases as the level of photometric uncertainty is increased from 1\%, to 5\% and then finally to more realistic HUDF levels. \label{paper10mconf}}
\end{figure}

We begin with a very simple case: noiseless observations of a $10^{10} M_{\sun}$ galaxy at a redshift of 4.5. 
This galaxy has a single  0.6 Gyr old stellar population, a global extinction of \AV=0.2 and a stellar metallicity 
of Z=0.001. We tested \PMCMC\ with many test objects with ages ranging from  very young to  old, with little to high 
extinction, and with low to high metallicities.  In this paper, we chose this example because these types of 
objects are not easily modeled using conventional SED fitting routines. The aim of this Section is to demonstrate 
how a limited set of photometric bands, combined with larger error bars, result in a significant broadening of the 
credible regions of the physical characteristic of an object. This is an expected result as it is intuitively obvious 
that as error bars are widened, a larger number of models become consistent with the observations. Furthermore, the
inter dependency and degeneracy between model parameters also leads to a widening of the credible regions. We 
illustrate this by showing the results obtained from using error bars which are 1\% of the flux, 5\% of the flux
and finally error bars which are the same size as those in the Hubble Ultra-Deep Field \citep[HUDF][]{Beckwith2006}.\\
\indent Figure \ref{1percent} shows several diagnostic plots. First, in panel (a), we plot the models that 
are contained in the MCMC chain. As we outlined earlier, an MCMC chain will spend more time in higher likelihood 
regions.  Likelier models will appear more often in the MCMC chain. This plot clearly shows that most 
of the models clearly fit within the photometric error bars.  As already mentioned in Section  \ref{convergence}, 
MCMC provides a formalism to estimate the errors associated with a posterior estimate of each given model parameter, 
but does not guaranty that a true likelihood maximum will be found within a preset number of iterations. Indeed, chains 
that are too short and have not converged to the true maximum likelihood it will lead to erroneous parameter estimates. 
In panel (b), we show the maximum value of the likelihood as a function of model parameter (marginalizing all others). 
This likelihood function is well behaved, smooth, and clearly peaked. Finally, panel (c) shows  the probability  
distribution function for each model parameter. From these, we derive our 95\% credible region and can state, 
(with a 95\% confidence), that this object is a Log(mass) of $10.^{0.01}_{-0.01}$ galaxy with a $0.61^{0.07}_{-0.06}$ 
Gyr stellar population and an extinction of $0.19^{0.08}_{-0.10}$.  While this is in excellent agreement with the 
fiducial values listed at the beginning of this Section, it appears that some model parameters are better contained 
than others (e.g Age and Mass versus \AV). \\
\indent  In Figure \ref{5percent} we show the effect of increasing the size of the error bars to 5\%. 
Everything else remains the same.  The figure demonstrates that increasing the size of the error bars flattens 
the shape of the maximum likelihood functions shown in Panel (b), particularly for extinction. 
Panel (c) shows much broader PDF for each parameter. It can only be stated 
with confidence that this object has:  Log(mass) $=$ $10. ^{0.04}_{-0.04}$; an age of $0.58 ^{0.19}_{-0.21}$ Gyrs
and an extinction of $0.23 ^{0.25}_{-0.23}$.  This is a significant broadening of the credible regions. 
It is particularly important to note that the PDF for \AV\ and Age are now very non gaussian.\\
\indent The two previous tests were performed with error bars that are significantly smaller than those seen in
deep surveys (e.g. HUDF and GRAPES).  In Figure \ref{1hudf} we now consider a more realistic case where the size of 
the error bars are set to  5\% for the ACS bands, 10\% for the WFC3 bands, and 15\% for IRAC1 and 20\% for the IRAC2. 
These values approximate the accuracy level of the data available for the HUDF. As panels (b) and (c) show, 
these realistic error levels lead to much more complex maximum likelihood functions with multiple peaks and PDFs that 
are very non gaussian. We now estimate that the galaxy has an age of $0.42 ^{0.30}_{-0.41}$ Gyr, an extinction 
\AV\ of $0.48 ^{1.6}_{-0.48}$, and a mass of  $10. ^{0.15}_{-0.19}$. The larger errors allow for two distinct models: 
young and dusty ($<0.2$Gyr old, \AV$>1.0$) and the fiducial one ($>0.2$Gyr old, \AV$<$0.6). The combination of model 
degeneracy and large error bars strongly limit one's ability to derive physical parameters using SED fitting. 
The increase in model degeneracy as the size of the error bars become larger is shown in Figure \ref{paper10mconf}. 
In this Figure we plot two dimensional probability densities where the degeneracy between stellar age and extinction are clear. 
As one goes from using 1\% error bars (top left panel) to 5\% (top right panel) to HUDF level error bars (bottom right panel), 
the 95\% credible interval (shown in blue) first stretches out and finally splits into multiple sub regions of acceptable parameters. 
Note that even the interval of acceptable stellar mass increases as different stellar populations with different mass to light ratios
become acceptable.

\subsection{More complex Cases: Increasing the Number of Free Parameters}\label{simple2}
\indent We have shown in Section \ref{simple} that model degeneracy limits our ability to constrain the physical characteristics 
of even the most basic galaxies. We now test \PMCMC\ with increasingly complex models. As noted in Section \ref{intro}, 
one advantage of MCMC is that it provides a computationally efficient way to compare observations to arbitrarily complex 
models. But, as we consider more complex models, the degeneracy between the various model parameters can only lead to wider 
posterior credible intervals. We first begin with metallicity in Section \ref{metal}, then add a second stellar population 
and nebular emission to create more complex models in Section \ref{complex}.

\subsubsection{Adding metallicity}\label{metal}
\indent We now add metallicity to the earlier parameters of mass, age, and extinction and test how \PMCMC\
handles this using the simulated observations described in Section \ref{simple}. Using a 1\% photometric uncertainty 
across all bands, we derive the 95\% credible regions of $0.61^{0.06}_{-0.05} Gyr$, \AV=$0.19^{0.09}_{-0.07}$, Z=$0.001^{0.001}_{-0.0001}$, 
and Log(mass) of $10.^{0.01}_{-0.02}$. This is very similar to the earlier results in which metallicity remained fixed.
The only differences are a slightly wider stellar mass uncertainty and a slight trend towards lower stellar masses.
Increasing the uncertainties to the 5\% levels yield $0.52^{0.20}_{-0.38} Gyr$, \AV=$0.21^{0.51}_{-0.20}$, Z=$0.002^{0.01}_{-0.0001}$,
and Log(mass) of $10.^{0.06}_{-0.11}$. These credible regions are now approximately twice as wide as the ones derived in 
Section \ref{simple} using 5\% error bars.   Finally, using HUDF level errors we obtain $0.14^{0.34}_{-0.11} Gyr$, 
\AV=$0.79^{0.60}_{-0.79}$, Z=$0.01^{0.02}_{-0.01}$, and Log(mass) of $9.9^{0.16}_{-0.16}$. These credible regions are once again 
several times wider than those derived earlier using HUDF error bar sizes but holding metallicity fixed. 
Model degeneracy has a strong impact on the size of the posterior credible regions for all parameters and wrongly assuming 
an a priori metallicity for a galaxy may lead to credible intervals that are artificially small.

\subsubsection{Increasingly complex models}\label{complex}
\indent We conclude this Section by showing the effect upon credible intervals for eight increasingly complex input models. 
The simulated galaxy is a $10^{10}M_\odot$ galaxy with two stellar populations of age 0.1Gyr (Z=0.0001) and 50Myr (Z=0.02).  
99\% of the stellar population is old with \AV=0.2. These values and population ratios where chosen so that the 
light produced by the younger stellar population would not be the dominant contribution to the stellar light of the simulated
galaxy. Unlike the previous Section, we now only consider realistic HUDF sized error bars.
Figure \ref{paper800hists} shows the results of applying \PMCMC\ using eight different parameters for 
the same set of simulated observations.  Each row of Figure \ref{paper800hists} shows the credible regions 
produced by \PMCMC\ for a given model. We start with a simple stellar population fit (SSP), allowing only the stellar age and mass to vary 
(Top Row). We then allow for extinction to be varied (Row 2). In turn, we then add the escape fraction for the nebular emission, 
metallicity, a second, older stellar population (SSP2), and finally set redshift as a free parameter. 
This Figure illustrates how continually adding model parameters produces increasingly non-gaussian PDFs. This, in turn,
weakens the constraints on the models. However, some parameters, such as mass, extinction, old stellar population age 
and redshift can still be reasonably constrained. While others, such as nebular emission and metallicity cannot be constrained.
Instead they become ``nuisance'' parameters, which add little to the final results.  For example, the nebular 
emission cannot be constrained.  Nominally, nebular emission contributes relatively little to objects with 
ages $>$ 10Myr \citep{reines2010}.  
Although \PMCMC\ is unable to constrain this parameter it does not seriously affect the quality of the 
posterior credible intervals of other parameters.
 
\subsection{Impact of Real Photometric Errors}\label{noise}
\indent We now address the issue of real photometric noise and how it may affect the MCMC derived posterior credible regions. 
The tests performed in Section \ref{simple} used noiseless data.  The error bars were simply some percentage
of the flux in each filter, while the flux value itself never varied.  In the case of testing real photometric errors,
the simulated fluxes in each filter are taken to be within some range of their ``true'' value in each iteration of the model. 
For example, if 10 iterations of a model are run, each of the 10 iterations will have a different value, as determined
by a random gaussian variate, for the flux.  Each iteration will be within some percent of the ``true'' value.  
This is a more realistic approach, as repeated observations of the same object will always vary to some degree, 
and the observer never knows the ``true'' value.  In order for \PMCMC\ to be useful, we must check its behavior and 
the accuracy of the credible intervals it derives using more realistic test cases. 

\subsubsection{Testing the reliability of the posterior credible regions}\label{credible}
\indent For the posterior credible regions to be useful they must translate to a simple reality: a 95\% credible region should 
include the fiducial value nearly 95\% of the time. We tested this by generating 200 simulated observations of a 
$10^{10}M_\odot$ galaxy with a 0.6Gyr old stellar population and \AV=0.2. We included the effect of photometric noise 
(assuming our previously defined HUDF error levels) and produced a set of 200 slightly different input SEDs to be analyzed using 
\PMCMC.  Examining the 95\% posterior credible regions and comparing them to the fiducial model parameters we find that input 
mass of  $10^{10}M_\odot$ was in the range predicted by the posterior 95\% credible interval for stellar mass 183 out of 200 times, 
(92\%).  Similarly, the fiducial stellar age and extinction were contained in the corresponding posterior credible regions 
90\% and 94\% of the time, respectively. This clearly demonstrates that we can rely on \PMCMC\ posterior credible regions 
to constrain physical characteristics of galaxies.

\subsubsection{Photometric noise level and its effect on posterior credible regions}\label{simplenoise}
\indent We have shown in Section \ref{simple} that the size of error bars has an effect on the width and shape of the 
posterior credible regions. We now examine for two distinct cases how decreasing the signal
to noise (S/N) of observations can change the derived posterior credible regions. We examine two test cases, at redshift 4.75.
This redshift is typical for the sources in \citet{pirzkal2007}.  Both cases have a stellar mass of $10^{10}M_\odot$ and 
an extinction of \AV=0.2. The first case is a galaxy with a metallicity of $Z=0.001$, and a stellar population that is 0.60 Gyr.
The second case is a galaxy with a 50Myr stellar population.   We generated 10 iterations for each of these simulated galaxies
with six increasing level of random noise added (0.1$\cdot$HUDF, 0.5$\cdot$HUDF, 0.75$\cdot$HUDF, 0.85$\cdot$HUDF, 
1.0$\cdot$HUDF, and$\cdot$1.5 HUDF). 
The 95\% posterior credible regions we derived are shown in Figures \ref{11mass} and \ref{12mass}.\\
\indent At photometric error levels of 0.1 HUDF, we can confidently estimate the stellar masses in 
both models (Top left panels in Figures \ref{11mass} and \ref{12mass}). However, by the time photometric errors match those 
of the HUDF, the posterior credible mass intervals span a factor of 3.0,. As more errors are added to each photometric band, 
the possibility of deriving a biased credible interval also increases. Both Figures demonstrate that the posterior credible 
intervals do include the fiducial parameter values. However, one can also see in Figure \ref{11mass} that increasing the sizes 
of the photometric errors leads to a larger number of acceptable models with younger stellar populations. This is because 
an observed SED can be made redder from either the presence of an old stellar population or by increasing \AV. 
On the other hand, a very young, blue population with little or no extinction has a very distinct SED shape. 
It is difficult to match anything other than a young SED to this type of population.

\subsubsection{The Impact of Adding the 3.6$\micron$ and 4.5$\micron$ IRAC Channels}\label{IRAC}
\indent At the redshifts considered here, IRAC observations using Channel 1 (3.6$\micron$) and Channel 2 (4.5$\micron$)
are often the only means we have to probe the restframe optical colors of high redshift galaxies. It is at these 
wavelengths that we may hope to constrain the contribution of any old stellar population(s). In addition these 
wavelengths probe the red side of the Balmer break.  At z=4.5, these IRAC bands correspond to $\approx 0.6\micron$ 
and $\approx 0.8\micron$. The main question we attempt to answer here is whether current IRAC observations
provide enough information to constrain the stellar ages and extinction of these sources. \\
\indent As a first example, we re-fit  the single stellar population galaxy from section \ref{simplenoise} 
with a single stellar population of 0.6Gyr, \AV=0.2 and a mass of $10^{10} M\sun$.  We use realistic HUDF level 
photometric noise.  We test the recovery of these parameters with and without the two IRAC channels.
The HUDF error levels are 15\% for IRAC1 and 20\% for the IRAC2.  The results are shown in Figure \ref{paper10iehists}. 
The results do not differ significantly.  The addition of the two IRAC channels only marginally improves the estimates
of the stellar ages.  With and without IRAC data yields: ages of $0.67^{0.19}_{-0.25}$Gyr vs. $0.64^{0.24}_{-0.32}$Gyr, 
\AV of $0.12^{0.29}_{-0.12}$ vs. $0.17^{0.53}_{-0.17}$), and total Log(mass) of $10.0^{0.11}_{-0.12}$ vs. $10.0^{0.20}_{-0.18}$. 
In this simple case, the IRAC measurements do help narrow our estimate of \AV, separating the degeneracy between the 
red colors of the older stellar population and the effects of extinction. However, it does not improve constraints
on the stellar ages.\\
\indent Next, the addition of the two IRAC channels are tested when considering the case of two stellar populations (SSP2)
using HUDF levels of photometric noise.  \PMCMC\ produces credible intervals for the ages that are very broad. 
Including the IRAC photometry produces credible intervals of Log(mass)= $9.53_{-0.60}^{0.52}$, compared to 
Log(mass)=$9.08_{-0.72}^{0.78}$ without them.  This is only a marginal improvement.  Our testing indicated that  
the only scenario which yields tangible improvements in the credible intervals occurs when the photometric errors in all
bands are kept at 0.1$\cdot$HUDF (i.e. $\approx 1\%$ for ACS and WFC3 bands and better than 3\% 
in the IRAC bands).  This produces posterior credible intervals of $0.655_{-0.13}^{0.12}$Gyr  and $0.04_{-0.00}^{0.01}$Gyr for 
the old and young stellar population, respectively, a mass fraction of $99.13_{-1.69}^{0.16}$, \AV\ of $0.27_{-0.27}^{0.08}$ and a 
total stellar Log(mass) of $10.05_{-0.14}^{0.06}$.   However, it should be noted that, to date, such small photometric errors 
have yet to be achieved.  

\subsection{Models and redshifts}
\indent Until now we have restricted our simulations to a narrow range of redshifts because we have been interested in 
determining the physical properties of high redshift galaxies for which detailed spectroscopic observations are difficult with 
current instrumentation. However, for \PMCMC\ to be a robust technique, it should be capable of working within a much larger
redshift space.   We now apply \PMCMC\ to a set of simulations with redshifts ranging from 1 to 8.   The simulated
galaxy we use has a 0.1 Gyr single stellar population, with an extinction of \AV=0.2, and a mass of ${\rm 10^{10}M_\odot}$.
In this section, we test how \PMCMC\ deals with determining credible intervals for mass, age and extinction using
SSP and EXP stellar population models over a large redshift range.  At each redshift, we have simulated the same object 
with random photometric noise added five separate times. We then used \PMCMC\ to generate multiple MCMC chains, as outlined in Section \ref{convergence}.\\ 
\indent Figures \ref{zmass} and \ref{zAv} show the results for the 8 redshifts applied to the simulations.  
Within each redshift, we show the 95\% credible regions for the stellar mass, stellar age, or \AV\ when using an SSP model 
with the IRAC channels (left most shaded part of each redshift bin), an EXP model with the IRAC channels 
(darker shaded part of the redshift bins), and finally using an SSP model without the use of IRAC data.  
These figures also show the credible region median values using black circles. Figure 
\ref{zmass} shows the while the mass is properly contained within the 95\% SSP credible regions it does allow for 
lower mass solutions. The EXP models produce credible regions that are more centered on the fiducial values 
(shown using a solid line).  The credible regions derived using an SSP model without the use of IRAC data shows a 
widening of the credible region towards lower masses, especially for redshifts higher than 5. \\
\indent Figure \ref{zmass} shows that the absence of IRAC data at z $>$ 4 results in a significant broadening of the credible
regions we derive. Furthermore, this broadening is not symmetrical and  the median values of the stellar masses and of the stellar ages are are lower than the 
fiducial values.  This is because at high redshifts the IRAC bands become 
more and more important for constraining the Balmer break.  Beyond that, the number of filters which can be used decreases due to drop-outs.
This forces us to rely on fewer data points to constrain the models, allowing for a very wide range of stellar ages, masses, as well as of the other model parameters
to become acceptable. \\
\indent  The credible regions for the extinction and metallicities for these simulations are shown in Figure \ref{zAv}. This Figure further shows how metallicity is most cases poorly constrained, which is something that we already saw earlier in this paper. 

\section{Applying \PMCMC\ to Observations}\label{science}
\indent In this section we now apply \PMCMC\ to real observational data of high redshift objects.
For each object, we used the SSP, SSP2, EXP, Maximum M/L SSP \citep{dickinson2003}, and SPP+Nebular models and have derived 
credible regions for the stellar mass, stellar ages,  and extinction for each of these objects.  The SSP, SSP2 and EXP 
models were described earlier in this paper. The maximum M/L SSP model is a variation of the two stellar population model 
(SSP2) where the older stellar population is fixed to have the maximum possible age (set to be the age of the Universe) 
at the redshift of the source. These models are often used to determine the maximum fraction of an object's mass that can 
be in the form of an older stellar population. 

\subsection{Observations}\label{observations}
\indent Nine GRAPES sources are examined in this paper.  The 9 objects are taken from the GRAPES
sources described in \cite{pirzkal2007}. As noted earlier, the GRAPES sample are based on spectroscopic
detections of \lya\ emission lines using the slitless grism mode of the Advanced Camera for Surveys (ACS)
on HST.  The 9 objects were selected because they each have a clear Ly-$\alpha$ detection, which provides
an independent confirmation of redshift. Since the GRAPES field nearly overlaps with the Hubble Ultra Deep 
Field (HUDF), there are many ancillary data available for these objects.  These fields have been observed by nearly 
every modern observatory in the world.  This allows us to constrain SEDs over a wide wavelength range. 
In \citet{pirzkal2007}, we relied on shallow NICMOS and VLT observations to constrain the near-UV rest frame 
luminosity and the rest-frame near UV slope. These data have since been supplemented by deep near-IR WFC3 observations. 
These new observations provide better constraints than the old NICMOS measurements. Using \PMCMC\ combined with
these new observations allows us to examine nature of these objects in more detail. Table \ref{phottable} 
lists the measured magnitudes of the \lya\ sources 
which are detected in the 2009 WFC3 observations of the HUDF (ERS;GO-11359). Out of the 9 sources listed 
in \cite{pirzkal2007}, 2 of them fall outside of the field of view of WFC3 (ID 631 and 712). In this paper, we take all other 
measurements of these sources in the observed optical and infrared bands (i.e. ACS and IRAC) from \cite{pirzkal2007}. 
The near-IR data are from the new WFC3 data. We use NICMOS measurements to supplement
GRAPES sources not observed with WFC3. The WFC3 data reduction is described in \cite{McLure2009}.
The photometry was measured in the same manner and using the same detection maps as described in \cite{pirzkal2007}.
The remaining sources discussed in this paper are taken from the literature and are a sample of 18 high z candidates, 
selected using the filter drop-out method, supplemented by 6 z$>$6 gravitationally lensed candidates. The 18 sources at z$\approx$7 and z$\approx$8 are taken from  \citet{labbe2010} who stacked these 
observations prior to SED fitting them.  Here, we instead examine the individual sources that comprise the stacked data.  The remaining 6 lensed sources are high redshift candidates 
lensed by nearby galaxy clusters and are taken from  \citet{bradley2008,zheng2009,richard2011}.
Table \ref{phottable} lists all  the photometric measurements used in this paper. Individual measurements for the 18 \citet{labbe2010} sources can be found in Table 1 of the original paper.

\subsection{Results}
\indent All of the results discussed in this Section are plotted in Figures \ref{highzmassagesav} and \ref{highzzfesc} 
and all of the credible regions we derive are listed in Tables \ref{sspres} to \ref{nebres}.

\subsubsection{The GRAPES sample}
\indent The five models (SSP, SSP2, EXP, Maximum M/L SSP, and SPP+Nebular) have been applied to all of the 
GRAPES sources and are shown in Figures \ref{highzmassagesav} and \ref{highzzfesc}. 
Figure \ref{g44422} shows the 1D distributions of the model parameters for one of our sources (4442). 
Tables \ref{sspres} to \ref{mupres} list the 68\% and 95\% credible intervals for all of these sources.  
In all cases, the use of a double population model does not significantly improve 
the fit. It does not yield significantly higher maximum stellar masses nor comes close to constraining 
the age of any possible old stellar populations in these objects. As demonstrated earlier, constraining old 
stellar ages requires photometric precision beyond what has been achieved to date.  However, we can conclude from 
Table \ref{sspres} that these sources do appear to be less than 50Myr old and the single population ages are 
 constrained to be low. Table \ref{sspres} lists the maximum formation redshifts of sources assuming an SSP population and 
maximum stellar population ages derived using the listed 95\% credible intervals. All GRAPES sources could have formed 
at a redshift ${\rm z<6}$.
The allowed range of extinction values is somewhat broad, but we estimate that \AV$\leq 1$ in these objects. 
Our results show the masses are 10$^{7}$ or 10$^{8}$ M${\odot}$, although they can 
vary up to a factor of twenty, depending on the specific star formation history model we consider.  
These are sub-$M^*$ values \citep[where $M^* = 3\times10^{10} M_\odot$,][]{bell2003}. 
In all cases, the addition of deep WFC3 near-IR photometry and the use of a more powerful SED fitting technique 
has confirmed that the GRAPES object are consistent with a very young stellar population \citep{pirzkal2007}.
When fitting the GRAPES sample using and SSP model with nebular emission, we find that four of them 
(4442, 5183, 9040, 9340) appear to have {\it f}$_{\rm esc}$ $<$ 0.5 and that two sources (631,6139) have 
{\it f}$_{\rm esc}$ $>$ 0.5.  The rest (712,5225,9487) have very broad ranges of acceptable values of {\it f}$_{\rm esc}$ .  
It should be noted that in this case \PMCMC\ produces credible intervals that are not limited to 
either very low or very high {\it f}$_{\rm esc}$ values.  This is different than recent work by \citet{Schaerer2011} 
which used a non-MCMC method based on grid-based parameter fitting and concluded that {\it f}$_{\rm esc}$
must either be negligible or close to unity.

\subsubsection{WFC3 HUDF objects}
\indent Recently, \citet{labbe2010} have identified several possible high redshift sources which the authors
claim may represent the most distant objects identified to date in the Universe.  The redshifts are based on 
WFC3 F850LP and F098M dropouts noted in the WFC3 Early-Release Science observations of the HUDF.  
The SED fitting was carried out on stacked data of 36 putative z$\sim$7 sources (data for 15 of which are given in their
Table 1) and 3 putative z$\sim$8 sources (all listed in their Table 1).  The authors chose to stack the flux densities of 
the z$\sim$7 sources in three $\sim$1 mag bins and stack the flux densities of the three z$\sim$8 sources.
BC03 SEDs assuming a Salpeter Initial Mass Function \citep{Salpeter1955} from 0.1 to 100 M$_{\odot}$ were then 
fit to the stacked objects (see their Figures 1 and 3) using a $\chi$$^{2}$ fitting technique.  
The best-fitting SEDs for the z$\sim$7 objects yield Log(mass)$\sim$9-9.8 with Z$\sim$0.2-1, with constant
star-formation (CSF) and a stellar age of 0.7 Gyr Myr (see their Figure 1). The z$\sim$8 stacked object yielded 
log(Mass) $\sim$ 9.3, Z$\sim$0.2, with CSF and a stellar age of 0.3 Gyr.  These results are 
particularly surprising, given that the age of the Universe at these redshifts is $\sim$ 0.6 Gyr. 
However, it is unclear what conclusions can be drawn from fitting stacked SEDs \citep[as discussed in][]{Nilsson2011}.\\
\indent The results from \citet{labbe2010} provide an interesting test for the \PMCMC\ method.
We started by first assuming nothing about the redshift of these sources. Unlike the GRAPES sample, these objects 
have no spectroscopic observations which can be used to confirm their redshifts. Thus, redshift was left as 
a free parameter to test whether their SEDs might be consistent with low redshift interlopers.  Fluxes for the sources
were taken from Table 1 of \citet{labbe2010}.  We note that their Table 1 lists negative flux densities 
for some objects.  In this case, and when detections were within 1$\sigma$ of the reported errors, we used
upper limits for these bands.  As noted earlier upper limits to the flux are used as part of a penalized likelihood.
We also note that their {\it Y}-band filter is actually a composite of fluxes from the F098M and F105W filters.
For the putative z$\sim$7 targets we applied \PMCMC\ to the individual sources and not a stacked composite, as
their Table 1 provides data for only 15/36 sources used to create their three stacked objects.  We also applied
\PMCMC\ to the three z$\sim$8 sources individually.\\
\indent  The results are shown in Figures \ref{highzmassagesav} to \ref{labbezfesc}. In Figures \ref{highzmassagesav} 
and \ref{highzzfesc}, we show the combined credible regions for the z$\approx$7 and for the z$\approx$8 sources.
They are labled L7 and L8.  These were determined by combining the individual credible regions of the 15 sources at z$\sim$7 
(L7) and combining the 3 individual sources listed at z$\sim$8 (L8).
As shown in these two figures, the combined credible regions indicate that these objects, as a whole, have stellar masses 
marginally larger than those of the GRAPES sample, and have potentially much larger values of extinction. 
The range of acceptable metallicities and {\it f}$_{\rm esc}$ are large and relatively unconstrained.
Given the broad composite credible regions and the absenece of any evidence suggesting these objects
are similar to each other,  it is worth examining them individually. 
Figures \ref{labbemassagesav} and \ref{labbezfesc}  show the individual credible regions of the  z$\approx$7 and z$\approx$8 
sources. We have labeled these L7-1 through L7-15 and L8-1 through L8-3, respectively  \citep[corresponding to the 
objects from the top to the bottom of Table 1 in][]{labbe2010}. These sources show a significant variation in
stellar mass. The masses within the group vary by as much as 100, with
some objects in the 10$^{8}$ M$_{\odot}$ range, compared to others with with masses $>$ 10$^{10}$ M$_{\odot}$
(e.g. compare L7-06 with L7-01 and L7-07).  \\
\indent  Figure \ref{labbemassagesav} show the acceptable stellar ages also span a broad range of valuess. 
Although in every case  we can exclude ages larger than 50-100 Myr old. This is in stark contrast to 
the results of \citet{labbe2010}, which claim ages of 300 Myr and 700 Myr for the stacked z$\sim$7 and z$\sim$8
sources, respectively.  The middle panel of Figure  \ref{labbemassagesav} shows the values of $\tau$ for the EXP model. 
It demonstrates that only moderate values ($<$ 4Gyr) are likely. Very large values of $\tau$ mean that the star formation 
history of these objects remains constant. This is again different from  \citet{labbe2010}. There, the authors 
assume a CSF (constant star-formaion) rate for the models they fit to the observations.\\
\indent  The bottom panel of Figure \ref{labbemassagesav} shows that the extinction varies greatly between sources. 
We see that a few of these sources (e.g. L7-01 and L7-06) are consistent with very low values of \AV,
while others have significantly larger values (e.g. L7-11 Av $\sim$ 1.5), while others have very large credible
intervals, indicating we cannot constrain \AV\ well.  Finally, \ref{labbezfesc} shows that both metallicity
and {\it f}$_{\rm esc}$ are difficult to constrain.  The latter is not surprising, given that at ages 50-100 Myr,
the contribution from nebular emission should be negligible \citep[e.g][]{starburst99,reines2010}.
%It is unclear why \citet{labbe2010} note the possibility of significant nebular contribution to the observed fluxes
%if their derived stellar ages of 300-700 Myr are correct. 
 Overall, we have demonstrated that the \PMCMC\ analysis
is able to provide robust information about the ages and masses of the individual objects. While some parameters
are less reliably constrained, we are still able to derive useful information about the nature of these objects. 

\subsubsection{Lensed $z>6$ candidates}
\indent We have applied the set of five models (SSP, SSP2, EXP, max M/L SSP2, and SSP+Nebular) models to six 
high redshift candidates identified by \citet{bradley2008}, \citet{zheng2009} and \citet{richard2011}. 
The objects are: A1689-zD1 from  \citet{bradley2008}; A1703-iD1, A2218-iD1, CL0024-iD1, and CL0024-zD1 from
\citet{zheng2009}; and A383-iD1 from \citet{richard2011}.  All of these objects are reported to be galaxies 
amplified by gravitational lensing.  These objects are interesting because
they may provide important information on the nature of galaxies during the period of reionization.  The
lensing provided by massive foreground clusters amplifies their observed light and sizes making it possible 
to not only detect these systems, but probe their properties. All three papers use standard $\chi$$^{2}$ fitting
of BC03 templates to derive physical properties.  \citet{bradley2008} conclude their object lies
at z$\sim$7.6 with a mass of 10$^{9}$ M$_{\odot}$ and a stellar population with an age of 45-300 Myr with 
no extinction.  \citet{zheng2009} find a mass of $\sim$ 10$^{10}$ M$_{\odot}$ for A1703-iD1 and 
$\sim$ 10$^{9}$ M$_{\odot}$ for CL0024-iD1 and CL0024-zD1 along with stellar ages $\sim$ 40-80 Myr with no extinction
for all 3 sources.  They do not report results for A383-iD1.
Finally,  \citet{richard2011} report their object lies at z$\sim$6 with a stellar mass $\sim$ 10$^{9}$, 
and no extinction.  They derive a best-fit age of $\sim$ 800 Myr, which (as the authors note themselves)
is difficult to reconcile given that the age of the Universe at that redshift is $\sim$ 900 Myr.\\
\indent Figures \ref{highzmassagesav} and \ref{highzzfesc} show the results when these lenses objects
are analyzed using \PMCMC. All candidate objects
except A383-iD1 show mass estimates that are ${\rm \approx 10^{10}}$ regardless of the five models used. 
This is $\sim$ 2-5$\times$ larger than those obtained from the fits in \citet{bradley2008} and \citet{zheng2009} for 
A1703-iD1, CL0024-iD1 and CL0024-zD1. Although it is consistent with the masses estimated for 
CL0024-iD1, and CL0024-zD1  \citep{zheng2009}.  A383-iD1 is discrepant as different models produce different results.  
This object could have a mass of $\sim$ 10$^{9-10}$ M$_{\odot}$ with a young population, low extinction, and high {\it f}$_{\rm esc}$,
or it could have a mass $>$ 10$^{11}$ M$_{\odot}$ and a stellar age approaching the age of the Universe (at that redshift).
As a group, these lensed galaxies are significantly more massive than the GRAPES or \citet{labbe2010} objects. 
We observe that the stellar age intervals, shown in Figure \ref{highzmassagesav} 
for the SSP, EXP, and SSP+Nebular models, are very broad, as are the credible regions for {\it f}$_{\rm esc}$ 
shown in Figure \ref{highzzfesc}. This yields stellar age estimates for the 6 sources that are poorly constrained. 
Improved physical parameter estimates for these sources require significantly more accurate photometry. 
However, we show in Table \ref{sspres} that the formation redshifts (z$_{\rm f}$), based on the upper 95\% 
credible intervals for the stellar age, are consistently less than ${\rm z=8.0}$ for these sources.

\subsection{A Note About Template Models}
\indent Briefly, we discuss possible differences between the stellar population models of BC03 and M05.  It has been noted 
in the literature that these models may yield differences in derived stellar masses and ages (among other parameters), 
particularly at longer wavelengths \citep[][e.g.]{Maraston2006}.  We have tested possible differences using SSP Models for
a subsample of the GRAPES sources from Table 1.  These tests compared the parameters of age, \AV, and stellar mass.
Table \ref{GRAPESSSPMA05} shows the results of these comparisons. The only significant differences are in the derived
ages.  BC03 ages are $\sim$ 2$\times$ older that those from M05.  The \AV\ and stellar masses are consistent with each other 
within the confidence intervals shown.  It is likely that the reason these differences are less pronounced is that
the comparisons between BC03 and M05 are at rest-frame optical wavelengths at these redshifts.  At rest-frame wavelengths
longer than $\sim$ 1$\micron$ BC03 and M05 show the greatest differences in parameters such as age, \AV\ and stellar mass.\\
\indent We note that the M05 templates provide equally good fits to our observations and do not significantly alter the width of 
95\% credible regions we derive. The credible regions derived using the M05 models are therefore essentially identical in shape, 
but shifted slightly in age compared to those derived using BC03.  

\subsection{High Redshift Photometric Redshifts}
\indent As was mentioned in Section \ref{intro}, redshift is one of the parameters that can be varied when applying 
\PMCMC, or any SED fitting algorithm. The GRAPES sources were spectroscopically identified using strong Ly-$\alpha$\ emission.
These identifications serve as the basis for the redshifts listed in Table \ref{phottable}. However, the non-GRAPES
sources in Table \ref{phottable} were photometrically selected. As an exercise, 
we applied \PMCMC\ to all of the sources in \ref{phottable}, including the GRAPES sources, to see how well we could recover 
these claimed redshifts. The GRAPES sources serve as a control-sample for this test.  \\
\indent The results are shown in Figure \ref{zphot}.   The photometric redshift derived for the GRAPES sample 
are relatively good matches to the spectroscopic ones.
%HOW MANY OF THE REAL Z FROM REDSHIFTS FALL INTO THE 95%INTERVALS? MIGHT WANT TO STATE THIS
The probability distributions for the remaining objects are shown in Figure \ref{zphot}.
Most redshifts are well defined.  This is expected since the effect of the Lyman break is a very strong photometric feature.
The lensed galaxy CL0024-zD1, identified by \citet{zheng2009} as a z$\sim$6-7 galaxy, shows two possible solutions from
the \PMCMC\ analysis: $z>6$ or $z<2.5$.  In Figure \ref{zphot2}, we show SED solutions for CL0024-zD1 obtained at z=0.2, z=7.9 and z=8.3
with the photometric fluxes over-plotted.  These particular solutions are the best fits in the regions $z<2.0$, $7.8<z<8.1$ and $z>8$, respectively. 
This demonstrates how photometric redshifts can produce ambiguous results when using only a limited number of 
available filters. A larger number of observations, including deep K band observations,  
deeper observations in the ACS bands (e.g. F606W, F775W, F850LP), and possibly low resolution WFC3 spectra are needed to 
confirm that this source is not a lower redshift interloper.

\section{Conclusions and Future Work}\label{conc}
\indent We have introduced our implementation of a Markov Chain based, SED analysis package, \PMCMC. We have tested \PMCMC\ extensively,
first against simulated galaxies of known parameters, and then using real observations of putative high redshift sources.
This analysis has made it possible to ascertain the caveats of comparing observed high redshift SEDs to models.  
Our main findings are:\\
%POINTS 2 AND 3 WERE SIMILAR SO I COMBINED THEM
\noindent {\bf 1)} That a Bayesian based Monte Carlo Markov Chain  analysis of the SEDs of high redshift  sources provides
a robust picture of the nature of these sources. Unlike more traditional best-fit based results, it can determine 95\% credible 
intervals for {\it each} parameter, including which ones are degenerate and which ones remain largely unconstrained.\\
\noindent {\bf 2)} Stellar mass is the easiest parameter to constrain.   \\
\noindent {\bf 3)} Photometric noise and choice of different star-formation histories have an impact on derived stellar mass, age, 
metallicity and extinction. In the case of stellar ages it is not possible to estimate a minimum age for the 
sources, but it is possible to determine upper limits (i.e. objects are less than a certain age). Metallicity cannot be 
constrained with the current level of photometric errors. Extinction and nebular emission (via {\it f}$_{\rm esc}$) are
often not constrained.\\
\noindent {\bf 4)} The addition of IRAC data, which probes the rest-frame optical light, is not enough to provide strong constraints
for the stellar ages and extinction. Significantly improved photometric precision, on the order of 1$-$2\% is required to constrain these quantities.\\
\noindent {\bf 5)} We tested our ability to constrain physical properties of objects using several star formation histories 
(SSP, EXP, SSP2, Maximum M/L SSP2) for $1<z<8$. We find that as redshift increases our ability to derive physical parameters decreases strongly.\\
\noindent {\bf 6)} We have applied our method, \PMCMC, to 33 sources at redshifts ranging from $4<z<8$: nine  GRAPES objects \citep{pirzkal2007},  
z$\sim$7-8 objects  \citep{labbe2010}, and six lensed $z>6$ candidates  \citep{bradley2008,zheng2009,richard2011}. 
We find significant differences between \PMCMC\ results and those in \cite{labbe2010}, particularly for stellar ages. We find 
for 2/4 lensed galaxies from \citet{zheng2009} mass differences in which \PMCMC\ produces values 2-5$\times$ larger.  The other 2 are
consistent with \citet{zheng2009}. In the case of the lensed galaxy from \cite{bradley2008}, we find three different credible redshift ranges, 
and two different, yet credible age and mass ranges (young and low-mass or extremely old and massive). We also demonstrate for these
sources that there is no statistically compelling evidence that any formed at a redshift larger than ${\rm z=8}$. \\

\begin{figure}
%fig800
\includegraphics[width=8in,angle=90]{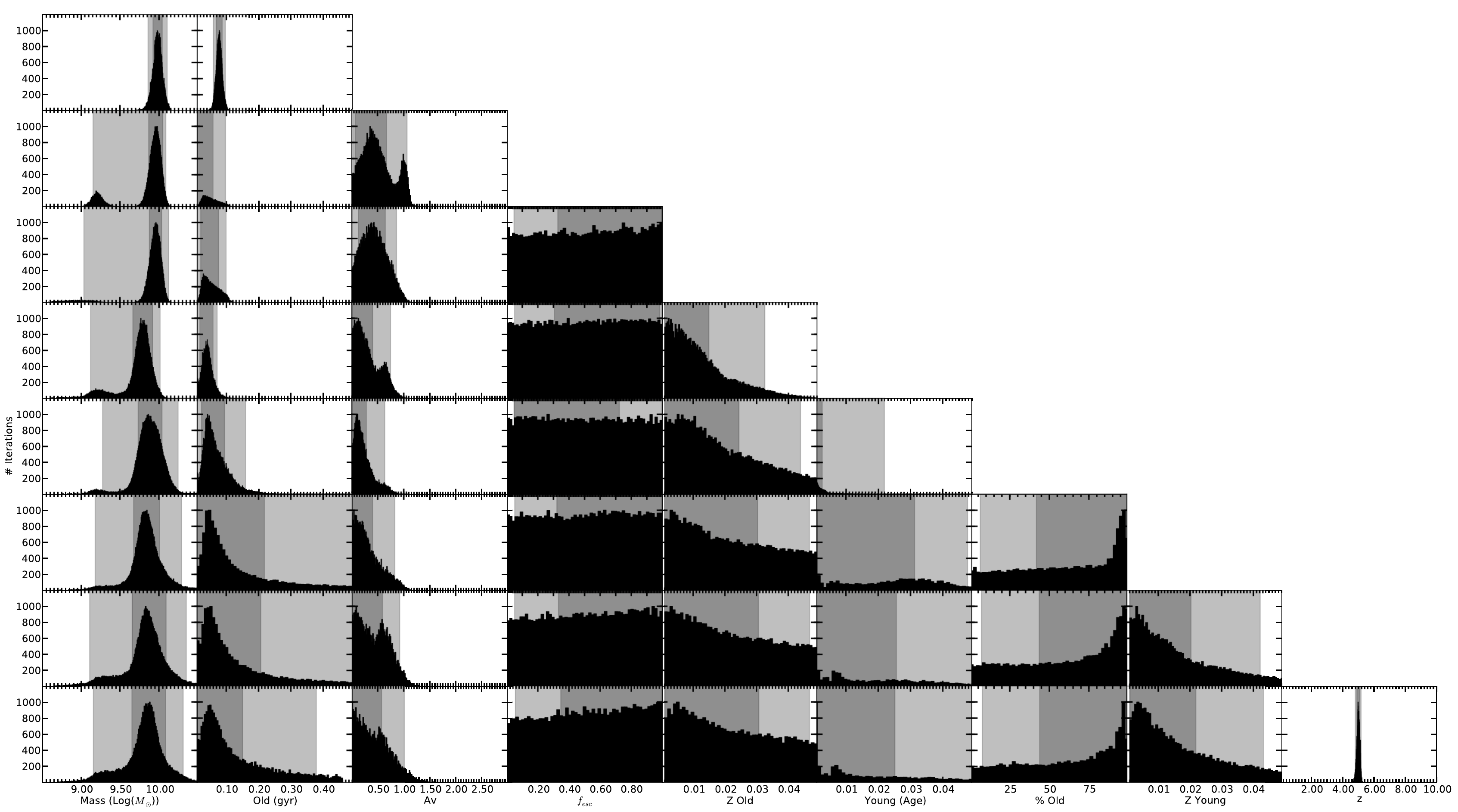} % fig800a_b.pdf
\caption{Model parameters resulting from apply \PMCMC\ using an increasingly complex model. We start (top row) by only allowing the mass and stellar population age to be varied. Each subsequent row then shows the result when adding another free parameter and running \PMCMC. As input, we have used a simulated galaxy at z=5 with 99\% of stars 0.1Gyr old with metallicity Z=0.001 and with 1\% of stars 50Myr old with metallicity Z=0.02 (solar). The extinction is \AV=0.2 and the total stellar mass is $10^{10}M_\odot$. \label{paper800hists}}
\end{figure}

\begin{figure}
%fig1011
\includegraphics[width=7.0in]{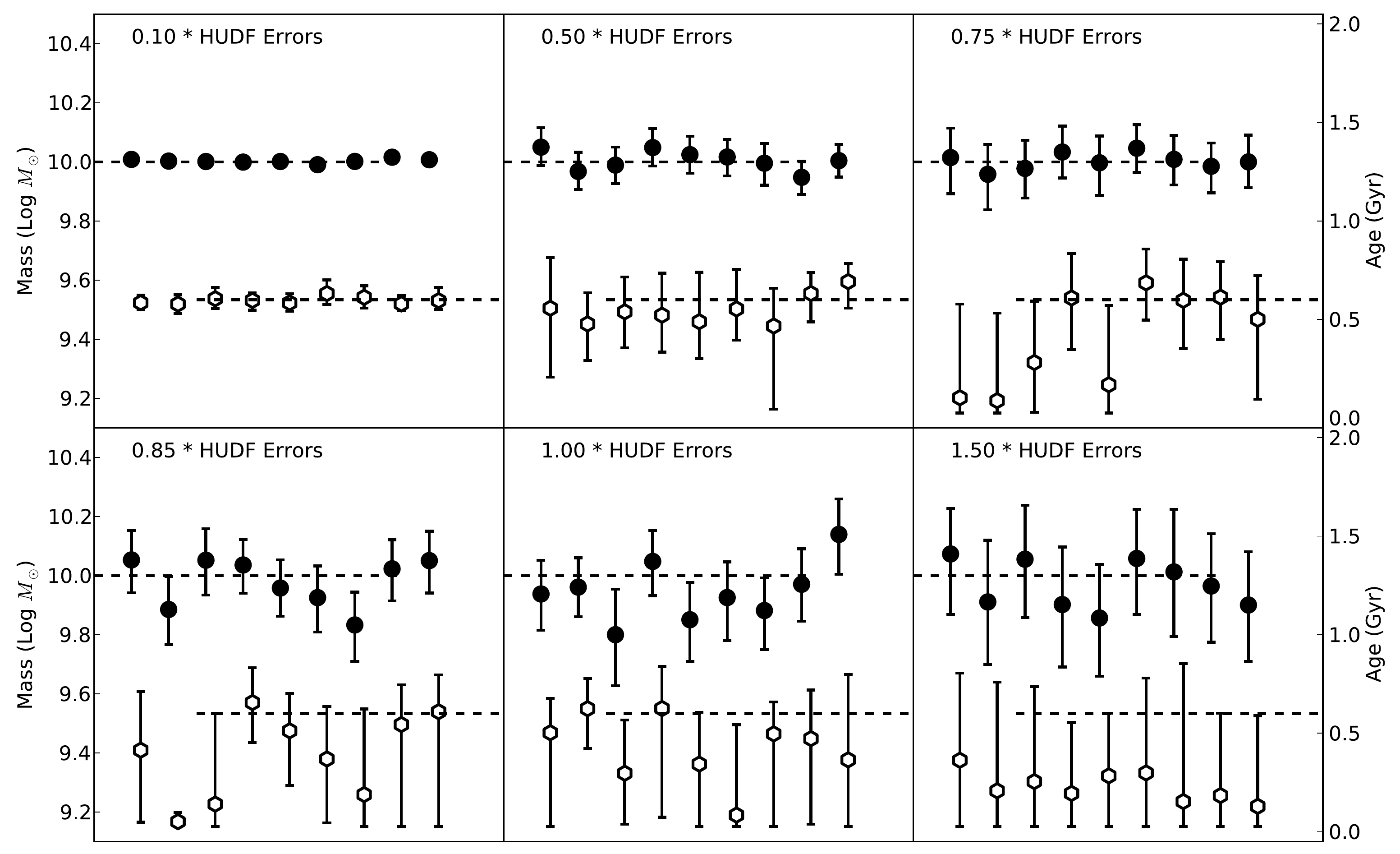}  %fig1011.pdf
\caption{Credible 95\% interval for the stellar population mass (dark circles) and stellar population age (white circles)  as a function of photometric error. Each panel shows ten independent simulations and the resulting 95\% credible region for the Log(mass) and stellar age of the object. Different panels show the results obtained under different photometric accuracy, ranging from 0.1 to 1.5 time that of typical high redshift HUDF error levels. The input model is a $10^{10} M_\sun$ galaxy with a single stellar population 0.6Gyr old and \AV=0.2, and is shown using dashed lines.\label{11mass}}
\end{figure}

\begin{figure}
%fig1213
\includegraphics[width=7.0in]{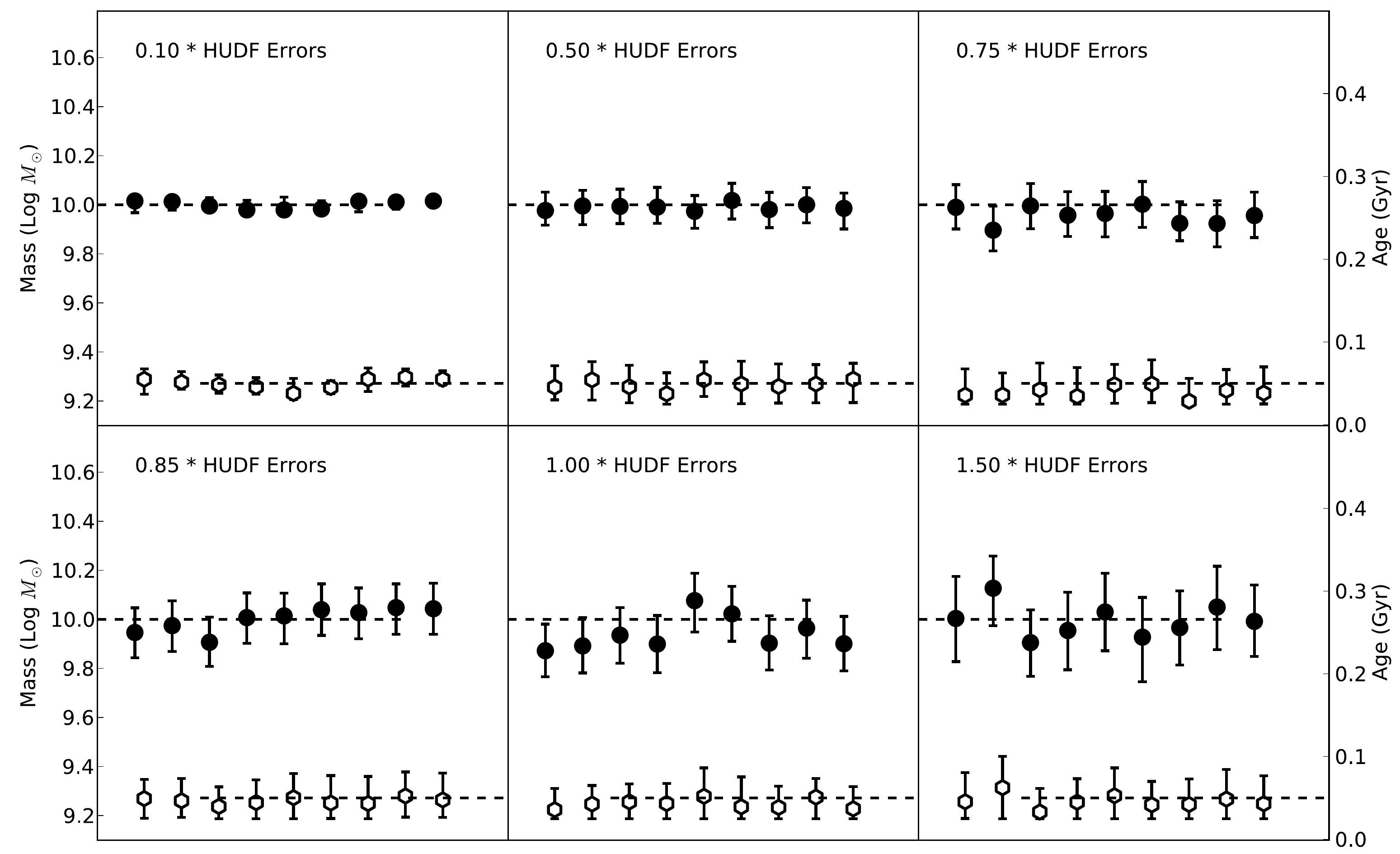} %fig1213
%%12mass.
\caption{Same as Figure \ref{11mass} but using input model that is a $10^{10} M_\sun$ galaxy with a single stellar population 50Myr old and \AV=0.2.\label{12mass}}
\end{figure}

\begin{figure}
%fig8
\includegraphics[width=7.0in]{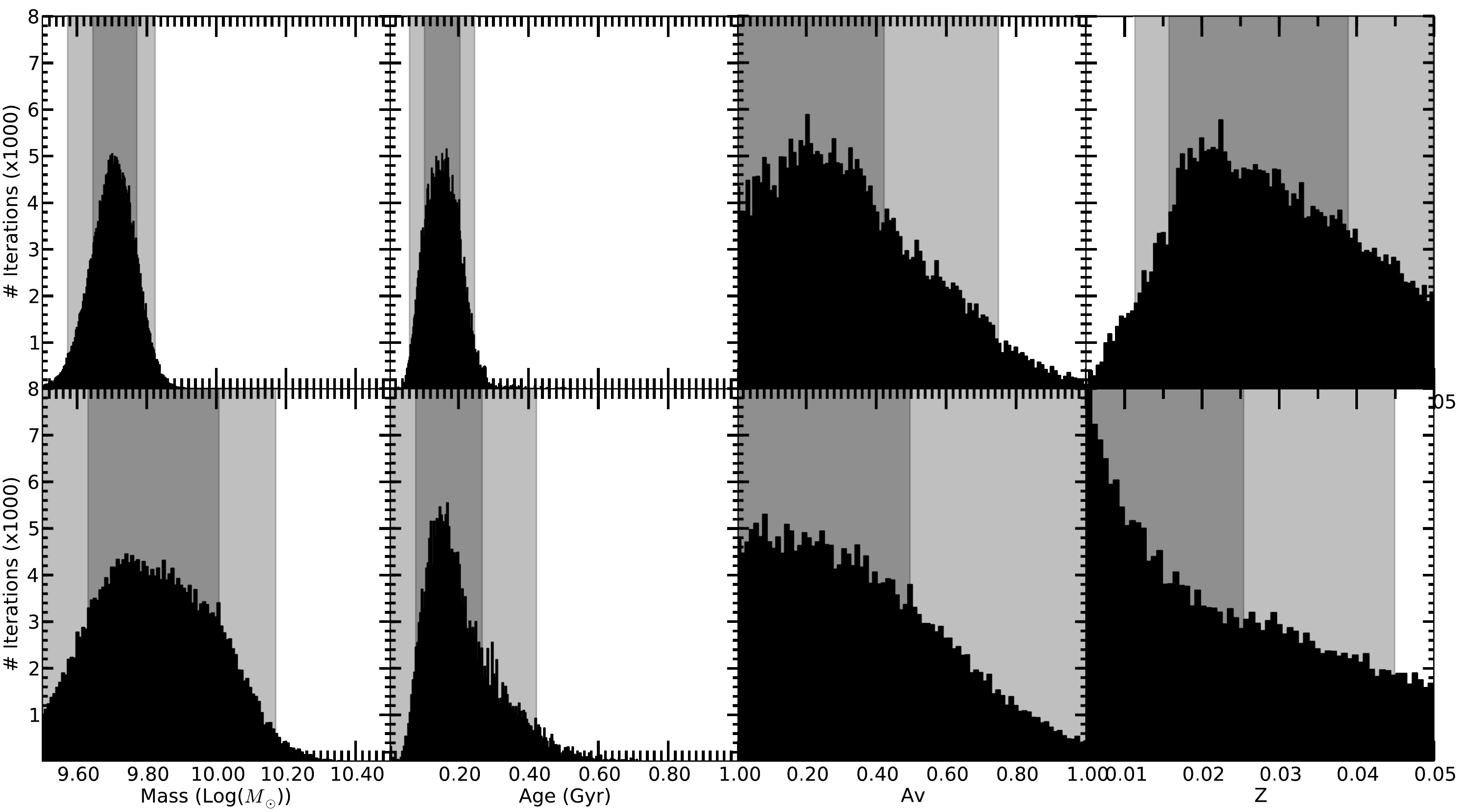} %fig8.pdf
\caption{Credible regions for the stellar mass, ages, extinction and metallicity for a simulated single population object at z=4.75 with (top row) and without (bottom row) IRAC observations probing the rest-frame optical light
of this object.
\label{paper10iehists}}
\end{figure}

\begin{figure}
%mass_z.pdf
\includegraphics[width =7in]{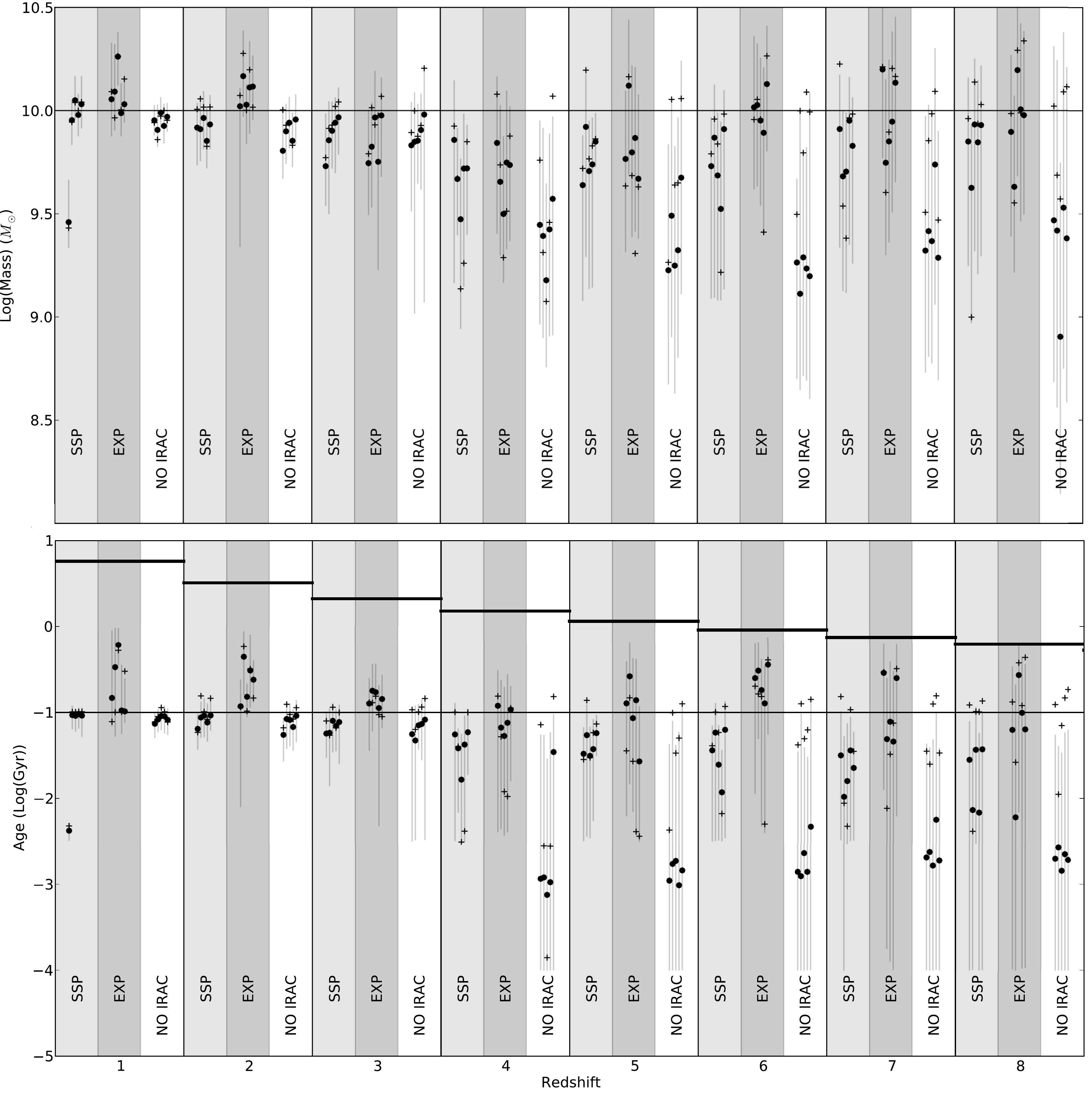} % z_mass_ages.pdf
\caption{Recovered 95\% credible regions for the stellar mass (top panel) and stellar ages (bottom panel) when applying \PMCMC\ to simulated observations of sources at redshifts ranging from z=1 to z=8. In each redshift band, we generated 5 random observations of each source, with realistic HUDF level photometric noise. We then use \PMCMC\ using a single stellar population model (SSP), and exponentially decaying SFH model (EXP) and repeated the SSP model but excluding the IRAC bands. At each redshift, the three applications of \PMCMC\ are shown with error bars. The best model fit values are shown with crosses. The median value of the derived credible intervals are shown using black circles. The true input model parameter (${\rm Log(mass)=10}$ and  ${\rm Log(Age)=-1}$) are shown using a thin horizontal lines. 
\label{zmass}}
\end{figure}

\begin{figure}
%mass_z.pdf
\includegraphics[width =7in]{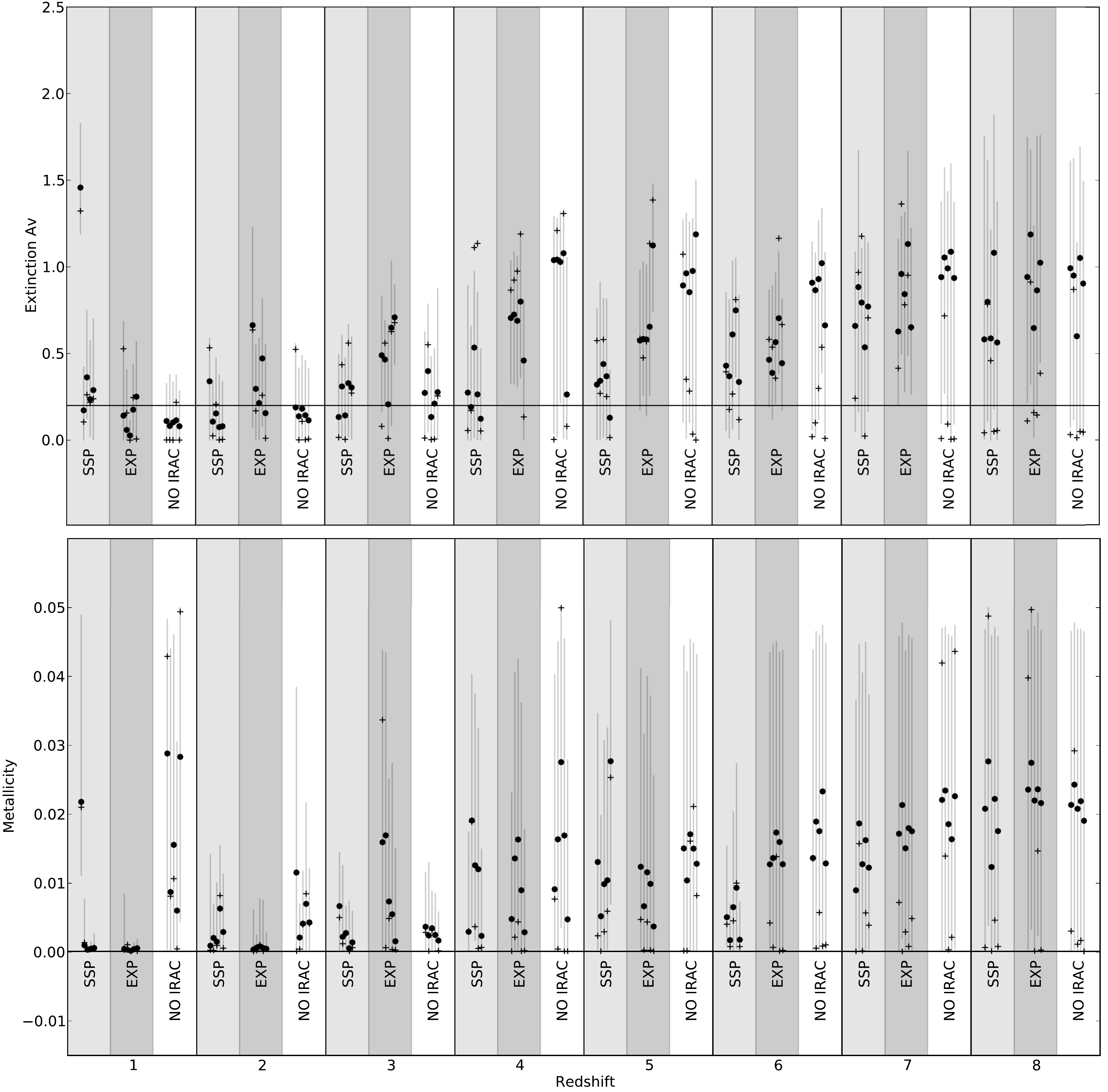} % z_Av_z
\caption{Recovered 95\% credible regions for the stellar extinction (top panel) and metallicity (bottom panel) and  when applying \PMCMC\ to simulated observations of sources at redshifts ranging from z=1 to z=8. In each redshift band, we generated 5 random observations of each source, with realistic HUDF level photometric noise. We then use \PMCMC\ using a single stellar population model (SSP), and exponentially decaying SFH model (EXP) and repeated the SSP model but excluding the IRAC bands. At each redshift, the three applications of \PMCMC\ are shown with error bars. The best model fit values are shown with crosses. The median value of the derived credible intervals are shown using black circles. The true input model parameter (\AV=0.2 and Z=0.001) are shown using a thin horizontal lines. 
\label{zAv}}
\end{figure}

\begin{figure}
%highz_mass
\includegraphics[width =5.5in]{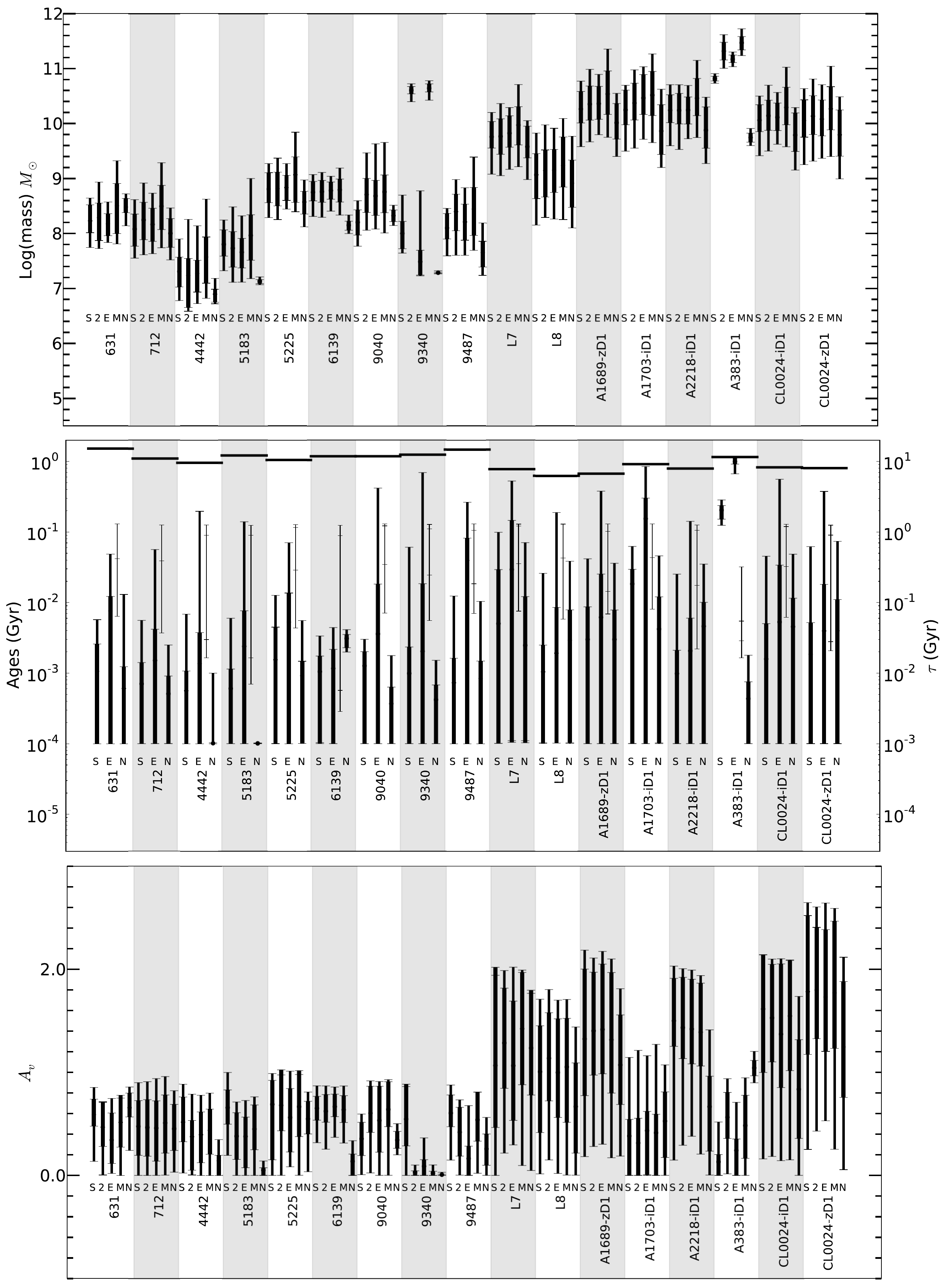} % highz_mass_ages_Av
\caption{The Top panel shows the median masses (black circles) and 68\% and 95\% credible regions (error bars)  for all of the high-z sources considered in Section \ref{science} of this paper. The stellar population ages and the extinction credible regions are shown in the Middle and Bottom panels, respectively. For each source, from left to right, we show the results of the different models we considered. The SSP, SSP2, EXP, Max M/L, and SSP+Nebular results with S, 2, E, M, and N, respectively. In the middle panel, values of $\tau$ are shown with thin error bars.
\label{highzmassagesav}}
\end{figure}

\begin{figure}
%highz_Ages
\includegraphics[width =5.5in]{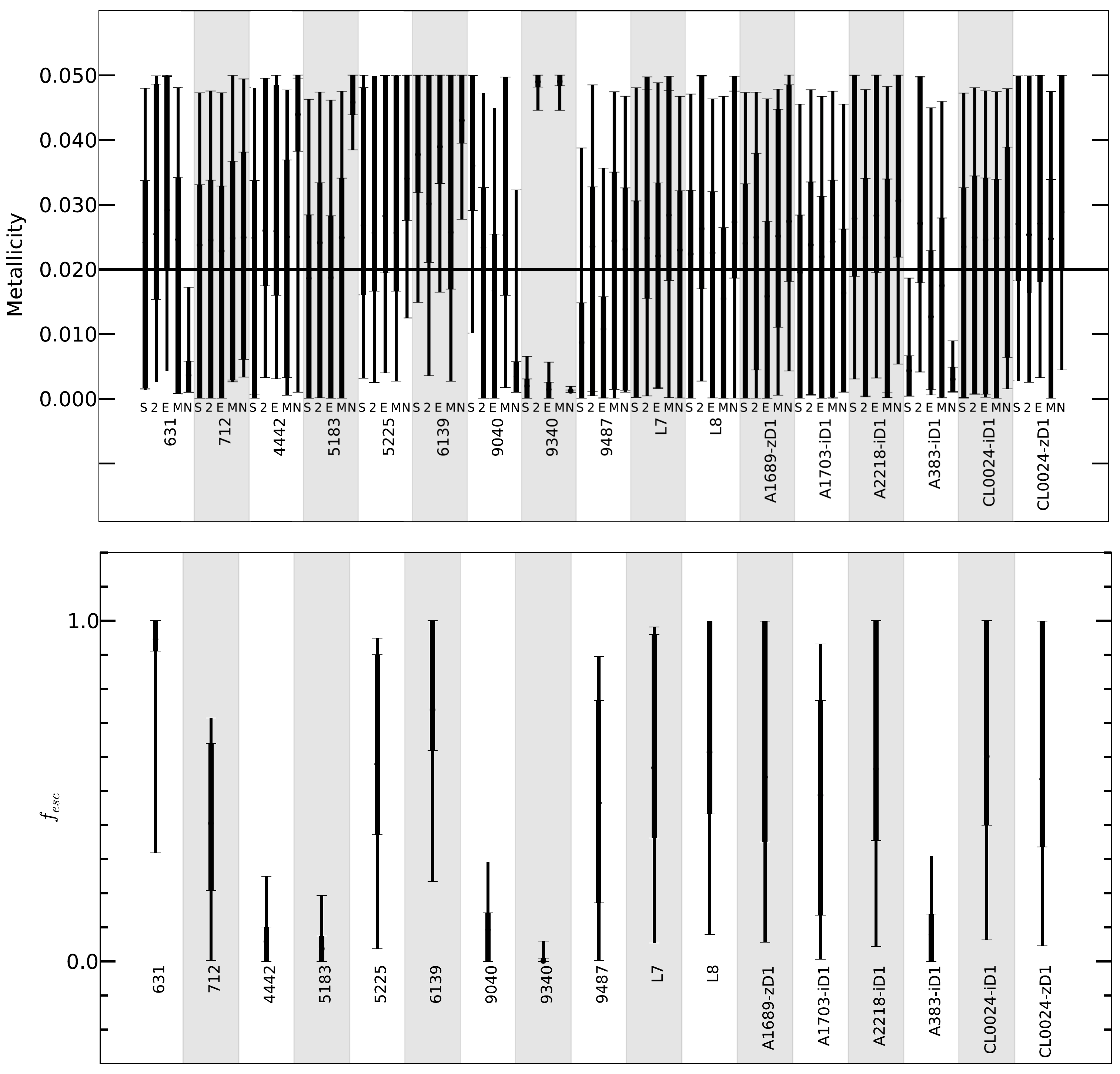} % highz_Z_fesc
\caption{This Figure is identical to Figure \ref{highzmassagesav}, we now shown the metallicity (top panel) and $f_{esc}$, the ionizing radiation escape fraction in the case of the SSP+Nebular mode (bottom panel).
\label{highzzfesc}}
\end{figure}

\begin{figure}
%fig23.pdf
\includegraphics[width =5.5in]{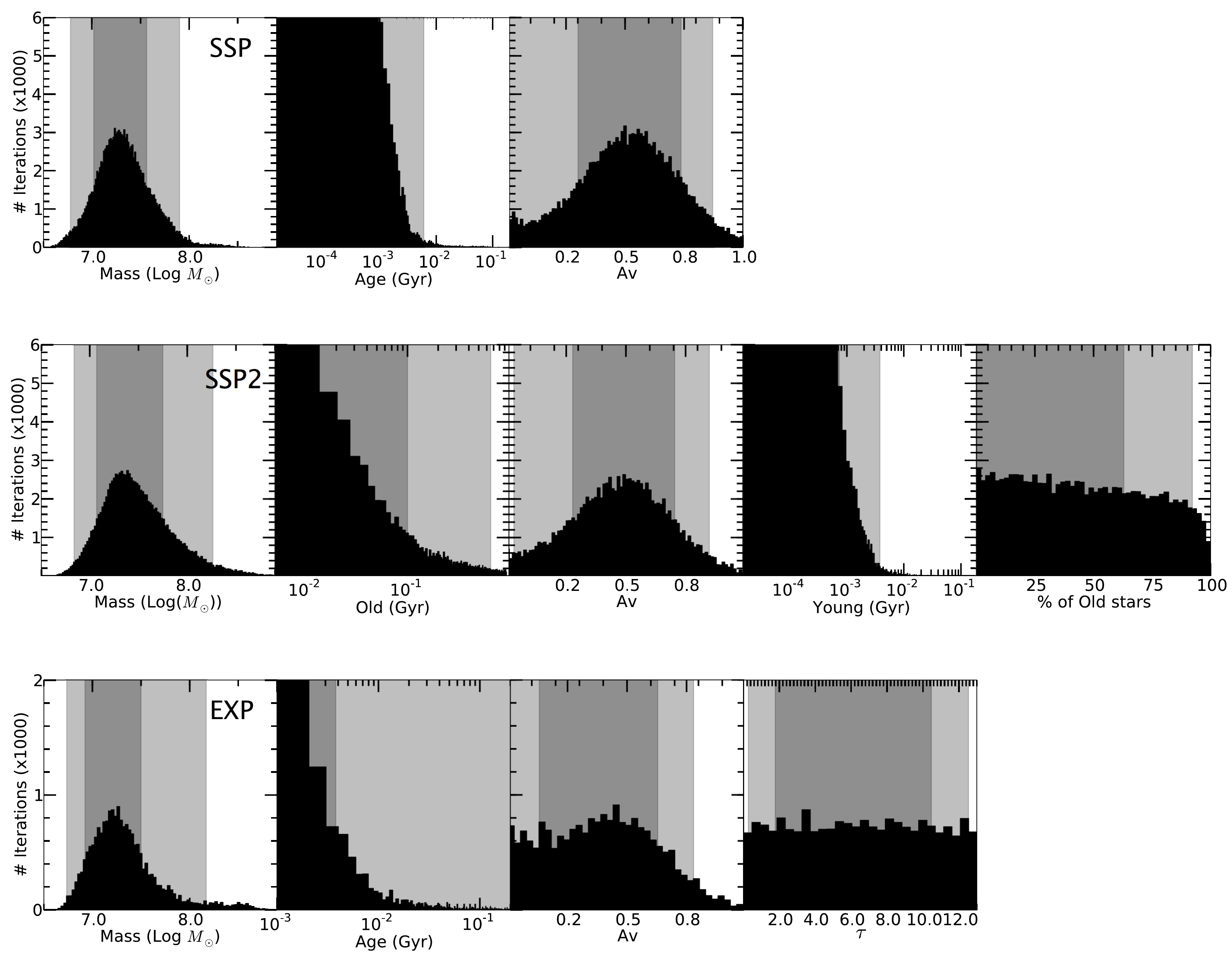} %g4442.pdf
\caption{Posterior probability densities of the model parameters of the SSP, SSP2, and EXP models when applied to the GRAPES 4442 object.
\label{g44422}}
\end{figure}

\begin{figure}
%highz_Ages
\includegraphics[width =5.5in]{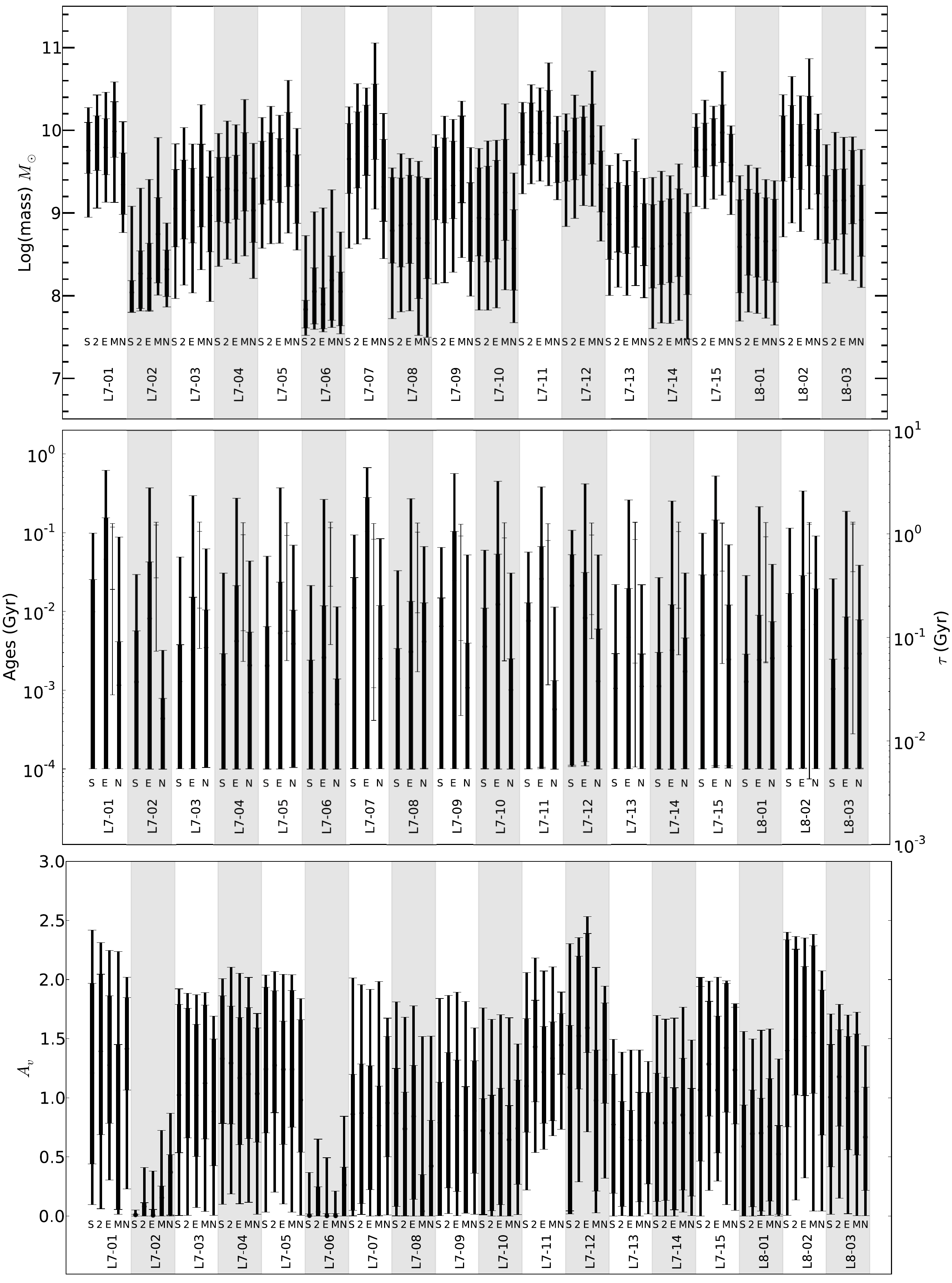} %labbe_mass_ages_Av.pdf
\caption{The 68\% and 95\% credible intervals for the individual sources listed in \citet{labbe2010}. The stellar mass is shown in the top panel. The stellar ages are shown in the middle panel. This panel also show the value of $\tau$ next to the EXP model stellar ages. The extinction is shown in the bottom panel. The SSP, SSP2, EXP, Maximum M/L, and SSP+Nebular models are shown left to right for each source and are labeled S, 2, E, M, and N, respectively.
\label{labbemassagesav}}
\end{figure}

\begin{figure}
%highz_Ages
\includegraphics[width =5.5in]{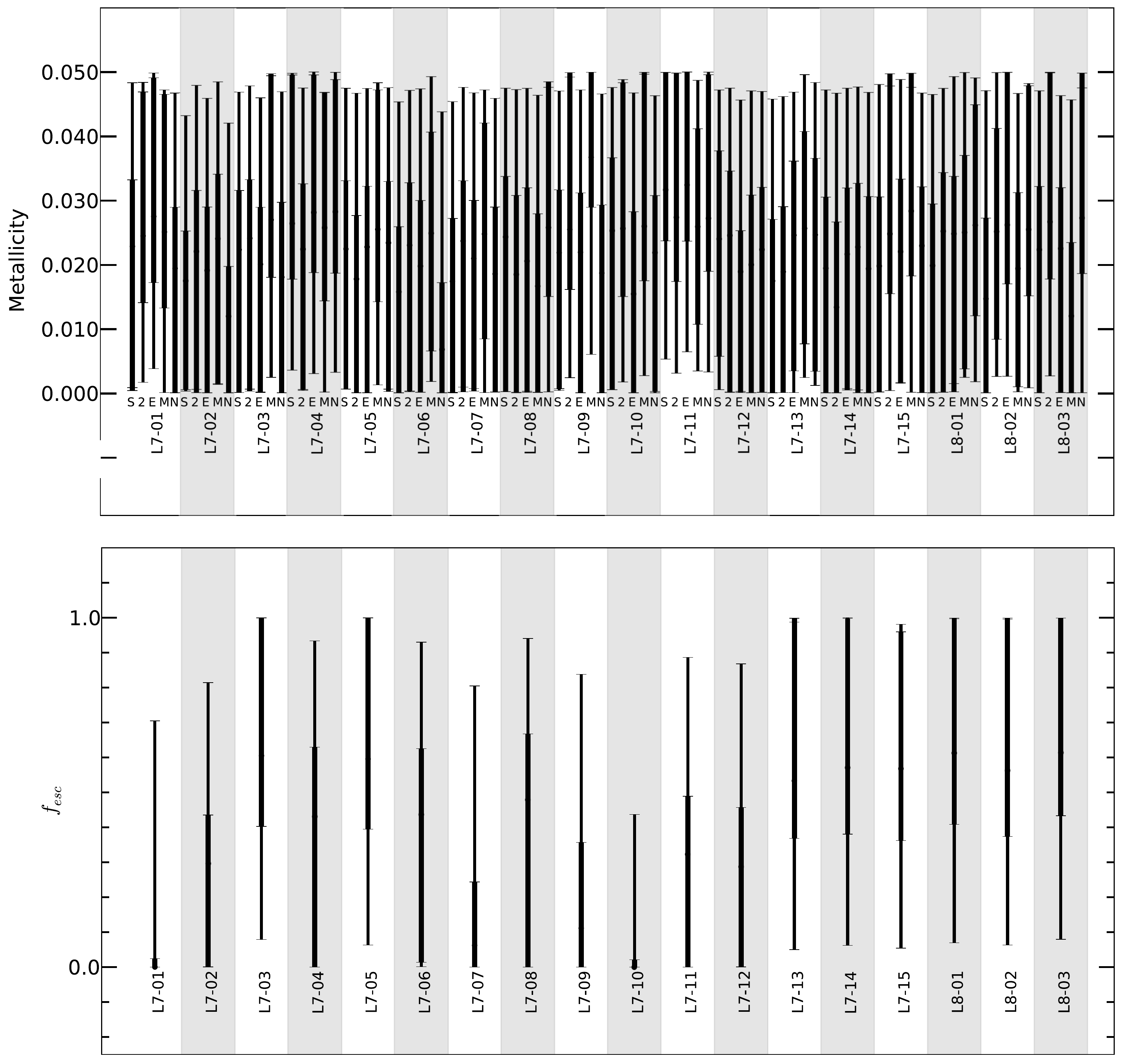} % labbe_Z_fesc.pdf
\caption{The 68\% and 95\% credible intervals for the individual sources listed in \citet{labbe2010}. The extinction, \AV,  is shown in the top panel. The nebular escape fractions are shown in the bottom panel.
\label{labbezfesc}}
\end{figure}

\begin{figure}
%photoz
\includegraphics[width =7in]{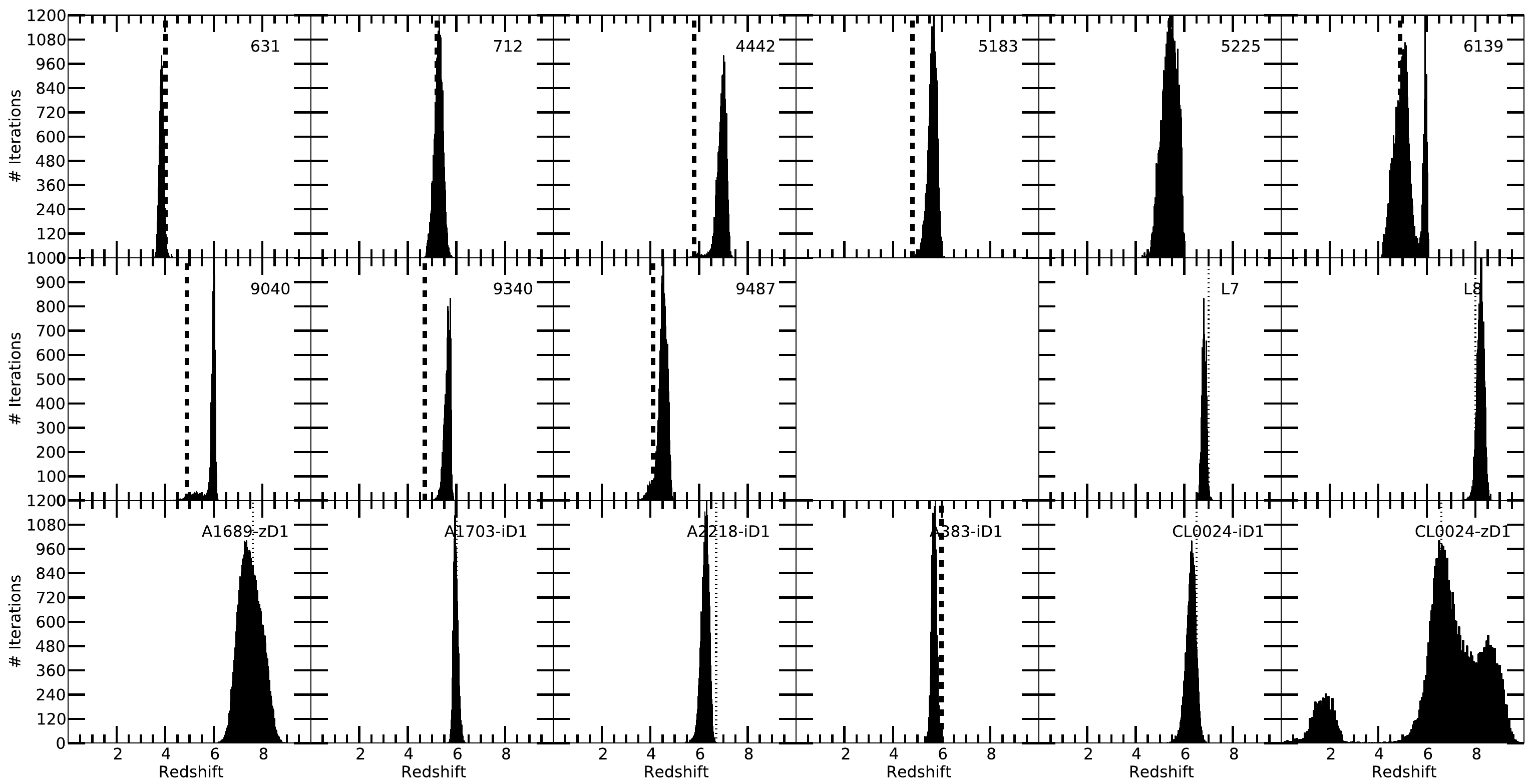} % photoz.pdf
\caption{The \PMCMC\ recovered probability distribution for the redshift of all seventeen sources listed in Table \ref{phottable}, using an SSP model and allowing redshift, stellar ages, mass and metallicity to vary. The spectroscopic redshifts are shown using thick dashed lines. Assumed photometric redshifts are shown using thin dotted lines.
\label{zphot}}
\end{figure}

\begin{figure}
%photoz2
\includegraphics[width =7in]{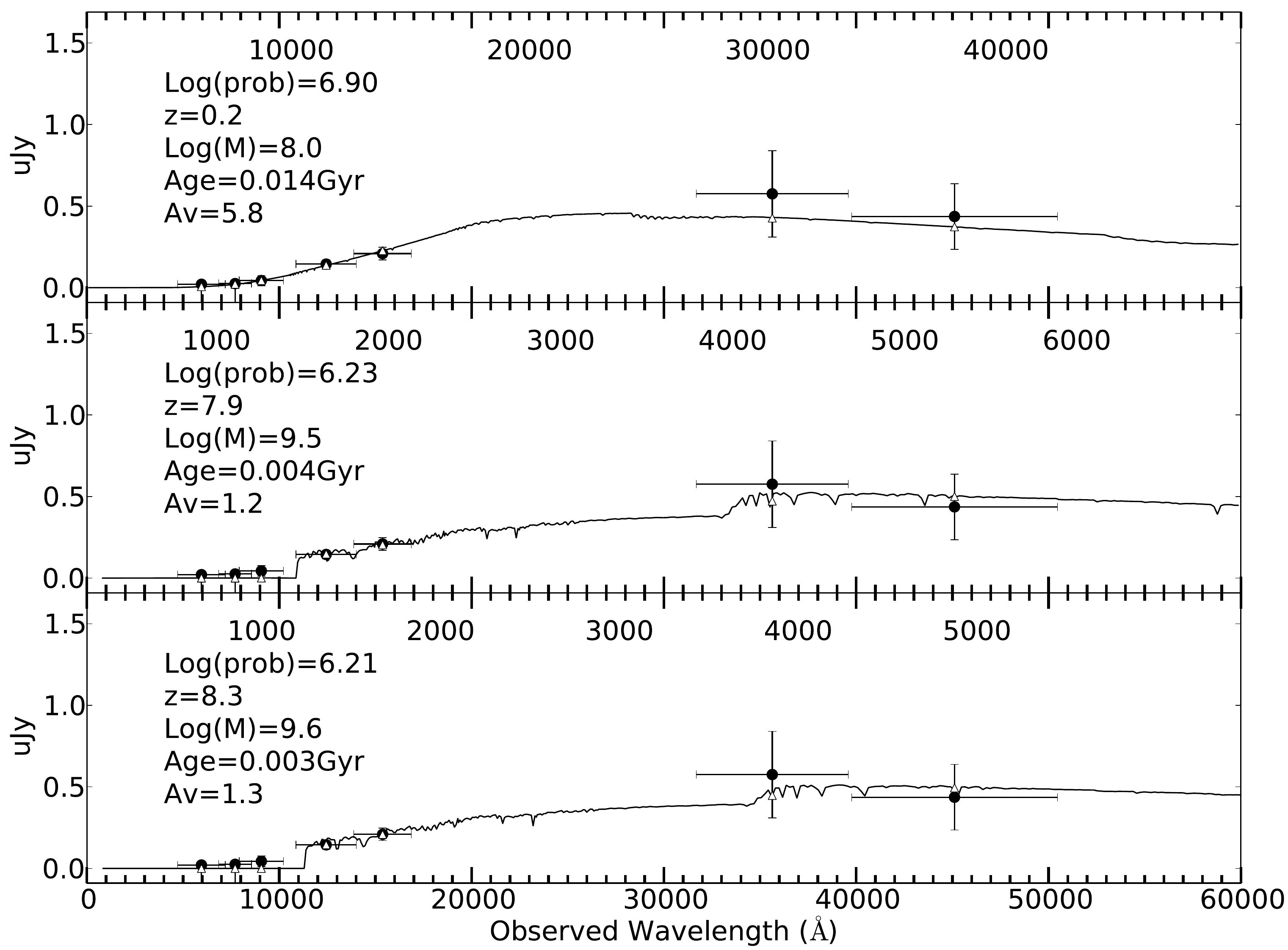} % photoz3.pdf
\caption{Possible solutions at $z<2.0$, $7.8<z<8.1$ and $z>8.1$ for object CL0024-zD1, shown in the top to bottom panels, respectively. Each panels shows the photometric redshift, and derived stellar mass, age, and extinction, as well as the Log-probability of the model that is over-plotted. the rest-frame wavelength is listed at the top of each sub-plot. The black circles and associated error bars show the observations. The white triangle show the model photometry in the same band passes.  This  illustrates  the difficulty in ruling out that high redshift sources such as this one, a confirmed lensed object at redshift  $\approx6$, are not low redshift interloper if one would only access to broad band photometry. 
\label{zphot2}}
\end{figure}

\indent We have shown that our implementation \PMCMC\ allows us to determine posterior credible intervals 
for the physical parameters of observed objects for a variety of input models, and with many possible parameters.
Our tests showed that high precision photometry is required to provide solid estimates of degenerate parameters such as age, extinction and metallicity.
The MCMC approach allows us to determine the range of possible model parameters in a manner that accounts for photometric errors as well as 
model degeneracies.   Overall, we have found that the \PMCMC\ approach to SED analysis is a powerful way to constrain physical parameters.
It allows one to fit observations and derive statistically credible intervals even when presented with complex models, degenerate parameters, 
or ``nuisance'' parameters.  Credible intervals are robust because they allow one to gauge the reliability (or believability) of each 
derived physical parameter.  \PMCMC deals with parameter degeneracies efficiently and allows us to easily identify the parameters 
that are the most important to the quality of the fit (e.g. Figure \ref{paper800hists}).
This is an attractive alternative to simple, best-fit methods as the latter cannot make an assessment of the
quality or reliability of each derived parameter, but rather yields a quality assessment of the entire fit.   \PMCMC\ and its Bayesian 
foundations let us identify a range of model parameter values that are consistent with the observations, assuming our model is correct.
However, one must remain conscious of the fact that we cannot be certain that the models which are compared to our observations are appropriate.
This is especially true at high redshifts where we have little evidence that stellar populations behave as they do in the local Universe.
Nevertheless, the exercises presented in this paper demonstrate that \PMCMC\ is a robust tool for deriving information for high redshift
sources.\\
\indent  Finally, we note that \PMCMC\ is not limited to broad-band photometry alone.  It can also be applied to low resolution spectra 
({\it R} $\sim$ 10s-1000).  Given the recent, successful deployment of the WFC3 grism during SMOV4, \PMCMC\ is an attractive tool for analyzing such data.
Furthermore, \PMCMC\ is not limited to the high-redshift universe, and can be applied to the constraining physical parameters 
of (relatively) nearby galaxies using low-resolution spectra.  Future work will include the analysis of optical and near-IR
spectra of coalesced galaxy mergers, including Luminous and Ultraluminous Infrared Galaxies \citep[e.g.][]{2010ApJ...712..318R}.

\acknowledgments

%===============================================================================
%%TABLES

\clearpage
\pagestyle{empty}
\begin{deluxetable}{lcccccccccc}
\tabletypesize{\footnotesize}
\setlength{\tabcolsep}{0.05in}
\tablewidth{0pt}
%\tablenum{1}
\pagestyle{empty}
\rotate
\tablecaption{Photometry\label{phottable}}
\tablecolumns{11}
\tablehead{
\colhead{UID}  &
\colhead{z}    &
\colhead{ACS}  & 
\colhead{ACS}  & 
\colhead{ACS} & 
\colhead{ACS}   & 
\colhead{WFC3} & 
\colhead{WFC3} & 
\colhead{WFC3}  &  
\colhead{IRAC1} & 
\colhead{IRAC2}  \\
\colhead{}  &
\colhead{}  &
\colhead{F435W}  & 
\colhead{F606W}  & 
\colhead{F775W} & 
\colhead{F850LP}   & 
\colhead{F105W} & 
\colhead{F125W} & 
\colhead{F160W}  &  
\colhead{} & 
\colhead{}
}
\startdata
631   &4.0  &29.59 $\pm$0.35  &26.87 $\pm$0.02  &26.78 $\pm$0.38       &26.62 $\pm$0.18  &26.49 $\pm$0.07\tablenotemark{a}    &\nodata          &26.76 $\pm$0.10\tablenotemark{a} &27.17 $\pm$0.37 &27.16 $\pm$0.44\\
712   &5.2  &31.48 $\pm$2.01  &29.66 $\pm$0.25  &28.06 $\pm$0.84       &27.12 $\pm$0.04  &\nodata                             &\nodata          &\nodata                          &28.05 $\pm$0.85 &$>$27.45 \\
4442  &5.8  &$>$32.96         &$>$31.64         &$>$31.40              &$>$29.82         &29.35 $\pm$0.06                     &29.16 $\pm$0.05  &29.45 $\pm$0.05                  &$>$28.53        &$>$28.07 \\
5183  &4.8  &$>$31.76         &$>$31.40         &28.61 $\pm$1.25       &27.97 $\pm$0.09  &28.18 $\pm$0.05                     &28.15 $\pm$0.05  &28.34 $\pm$0.05                  &$>$28.13        &$>$27.85 \\
5225  &5.4  &$>$28.88         &$>$28.29         &$>$26.21              &25.93 $\pm$0.04  &25.94 $\pm$0.05                     &25.87 $\pm$0.05  &25.95 $\pm$0.05                  &26.51 $\pm$0.28 &25.86 $\pm$0.26\\
6139  &4.9  &$>$32.92         &$>$26.97         &25.96 $\pm$0.45       &25.66 $\pm$0.02  &25.56 $\pm$0.05                     &25.57 $\pm$0.05  &25.64 $\pm$0.05                  &26.35 $\pm$0.23 &27.10 $\pm$0.39\\
9040  &4.9  &$>$30.14         &$>$28.33         &$>$26.17              &26.23 $\pm$0.05  &26.15 $\pm$0.05                     &26.28 $\pm$0.05  &26.31 $\pm$0.05                  &$>$28.45        &$>$28.36 \\
9340  &4.7  &$>$33.04         &$>$31.40         &28.25 $\pm$0.77       &27.69 $\pm$0.07  &27.86 $\pm$0.05                     &28.61 $\pm$0.05  &28.02 $\pm$0.05                  &$>$27.48        &$>$27.16 \\
9487  &4.1  &$>$30.07         &$>$28.24         &27.16 $\pm$0.03       &27.27 $\pm$0.05  &27.13 $\pm$0.05                     &27.41 $\pm$0.05  &27.42 $\pm$0.05                  &27.47 $\pm$0.47 &$>$27.93 \\

L7-01\tablenotemark{1}   &7 &$ >29.89 $ & $ >31.20 $ & $ > 29.31 $ & $ 28.57 \pm 0.7$ &   $ 27.37 \pm 0.29$ & $ 27.18 \pm 0.18$ & $ 26.93 \pm 0.18$ & $ 25.52 \pm 0.25$ & $26.08 \pm 0.72$\\
L8-01\tablenotemark{1}   &8 & $ >29. $ & $ >30.50  $ & $> 29. $   & $ >29. $ & $ >30.89$ &    $ 27.38 \pm 0.21$ & $ 27.47 \pm 0.20$ & $ >27.71 $ & $ >26.92 $\\

A383-iD1\tablenotemark{2}   &6.  &\nodata &\nodata  &$>$26.48         &24.54 $\pm$0.12  &24.65 $\pm$0.05\tablenotemark{b}    &24.57 $\pm$0.08  &24.66 $\pm$0.05                  &23.06 $\pm$0.08 &22.83 $\pm$0.08\\
                   &   &\nodata &\nodata  &          &25.63 $\pm$0.14\tablenotemark{c} &             &          &                          &         &\\
A1689-zD1\tablenotemark{3}   &7.6  &\nodata &$>$27.80 &$>$27.80         &$>$27.50         &\nodata                             &25.30 $\pm$0.10\tablenotemark{d}  &24.70 $\pm$0.10\tablenotemark{d}                  &24.20 $\pm$0.30 &23.90 $\pm$0.30\\
A1703-iD1\tablenotemark{4}  &6.  &\nodata &$>$28.20 &26.80 $\pm$0.30  &24.20 $\pm$0.10  &\nodata                             &24.00 $\pm$0.10  &23.90 $\pm$0.10                  &23.10 $\pm$0.30 &23.50 $\pm$0.40\\
A2218-iD1\tablenotemark{4}  &6.7  &\nodata &$>$27.60 &$>$27.20         &25.10 $\pm$0.10  &\nodata                             &24.30 $\pm$0.10  &24.10 $\pm$0.10                  &23.70 $\pm$0.30 &23.90 $\pm$0.30\\
CL0024-iD1\tablenotemark{4} &6.5  &\nodata &$>$28.10 &$>$27.90         &26.00 $\pm$0.20  &\nodata                             &25.10 $\pm$0.10  &25.00 $\pm$0.10                  &24.40 $\pm$0.20 &24.40 $\pm$0.30\\
CL0024-zD1\tablenotemark{4} &6.6  &\nodata &$>$28.10 &$>$27.90         &27.30 $\pm$0.80  &\nodata                             &26.00 $\pm$0.20  &25.60 $\pm$0.20                  &24.50 $\pm$0.50 &24.80 $\pm$0.50\\
\enddata
\tablecomments{\\
Upper limits are $1\sigma$.\\
(a) HST/NICMOS measurements with F110W Filter\\
(b) WFC3 measurements with F110W Filter\\
(c) ACS/WFC measurements with F814W filter\\
(d) HST/NICMOS measurements with F110W and F160W Filter\\
(1) Sample object with photometric redshift from  \cite{labbe2010}\\
(2) Lensed object with possible spectroscopic confirmation from \cite{richard2011}\\
(3) Lensed object with photometric redshift from  \cite{bradley2008}\\
(4) Lensed object with photometric redshift from  \cite{zheng2009}\\
}
\end{deluxetable}

\clearpage

\clearpage
\pagestyle{empty}
\begin{deluxetable}{cccccc}
\tabletypesize{\footnotesize}
\setlength{\tabcolsep}{0.06in}
\tablewidth{0pt}\tablecaption{\PMCMC\ SSP results for object listes in Table \ref{phottable}\label{sspres}}
\tablehead{\colhead{UID} &  \colhead{Mass (${\rm M_\sun}$)} & \colhead{Age (Gyr)}  & \colhead{\AV\tablenotemark{a}} & \colhead{Z\tablenotemark{b}}   & \colhead{${\rm z_f}$\tablenotemark{c}} }
\startdata
631&$8.09^{+0.51,+0.24}_{-0.40,-0.31}$&$0.003^{+0.003,0.001}_{-0.003,-0.003}$&$0.49^{+0.27,+0.24}_{-0.49,-0.30}$&$0.02^{+0.02,+0.01}_{-0.02,-0.02}$&4.0\\
712&$8.16^{+0.57,+0.30}_{-0.58,-0.34}$&$0.001^{+0.008,0.001}_{-0.001,-0.001}$&$0.52^{+0.45,+0.29}_{-0.52,-0.29}$&$0.02^{+0.03,+0.01}_{-0.02,-0.02}$&5.2\\
4442&$7.24^{+0.58,+0.24}_{-0.49,-0.29}$&$0.001^{+0.006,0.001}_{-0.000,-0.000}$&$0.45^{+0.38,+0.22}_{-0.43,-0.21}$&$0.03^{+0.02,+0.02}_{-0.02,-0.01}$&5.8\\
5183&$7.58^{+0.50,+0.29}_{-0.57,-0.33}$&$0.001^{+0.012,0.001}_{-0.001,-0.001}$&$0.39^{+0.35,+0.12}_{-0.39,-0.39}$&$0.02^{+0.03,+0.01}_{-0.02,-0.02}$&4.8\\
5225&$8.75^{+0.49,+0.26}_{-0.49,-0.32}$&$0.005^{+0.010,0.002}_{-0.004,-0.004}$&$0.37^{+0.61,+0.38}_{-0.37,-0.37}$&$0.03^{+0.02,+0.02}_{-0.03,-0.01}$&5.5\\
6139&$8.76^{+0.31,+0.19}_{-0.42,-0.15}$&$0.001^{+0.002,0.001}_{-0.001,-0.001}$&$0.65^{+0.22,+0.13}_{-0.28,-0.09}$&$0.04^{+0.01,+0.01}_{-0.02,-0.01}$&4.9\\
9040&$8.62^{+0.50,+0.29}_{-0.58,-0.24}$&$0.001^{+0.010,0.001}_{-0.001,-0.001}$&$0.65^{+0.34,+0.24}_{-0.58,-0.17}$&$0.02^{+0.03,+0.01}_{-0.02,-0.02}$&4.9\\
9340&$7.47^{+0.43,+0.17}_{-0.41,-0.12}$&$0.001^{+0.009,0.001}_{-0.001,-0.001}$&$0.08^{+0.31,+0.07}_{-0.08,-0.08}$&$0.00^{+0.00,+0.00}_{-0.00,-0.00}$&4.7\\
9487&$8.06^{+0.44,+0.27}_{-0.51,-0.26}$&$0.001^{+0.019,0.002}_{-0.001,-0.001}$&$0.55^{+0.25,+0.29}_{-0.55,-0.17}$&$0.01^{+0.03,+0.00}_{-0.01,-0.01}$&4.1\\
L7-01&$9.75^{+0.52,+0.34}_{-0.80,-0.27}$&$0.010^{+0.089,0.015}_{-0.010,-0.010}$&$1.21^{+1.21,+0.76}_{-1.11,-0.77}$&$0.02^{+0.03,+0.01}_{-0.02,-0.02}$&7.8\\
L8-01&$8.59^{+0.86,+0.57}_{-0.90,-0.55}$&$0.001^{+0.027,0.002}_{-0.001,-0.001}$&$0.59^{+0.97,+0.35}_{-0.59,-0.59}$&$0.02^{+0.03,+0.01}_{-0.02,-0.02}$&8.3\\
A1689-zD1&$10.26^{+0.51,+0.32}_{-0.68,-0.25}$&$0.003^{+0.039,0.006}_{-0.003,-0.003}$&$1.32^{+0.86,+0.68}_{-1.14,-0.55}$&$0.02^{+0.02,+0.01}_{-0.02,-0.02}$&8.0\\
A1703-iD1&$10.25^{+0.45,+0.36}_{-0.75,-0.27}$&$0.018^{+0.044,0.011}_{-0.018,-0.018}$&$0.38^{+0.76,+0.16}_{-0.38,-0.38}$&$0.02^{+0.03,+0.01}_{-0.02,-0.02}$&6.3\\
A2218-iD1&$10.24^{+0.46,+0.28}_{-0.64,-0.23}$&$0.001^{+0.024,0.001}_{-0.001,-0.001}$&$1.50^{+0.53,+0.41}_{-1.35,-0.25}$&$0.03^{+0.02,+0.02}_{-0.02,-0.01}$&6.9\\
A383-iD1&$10.81^{+0.09,+0.05}_{-0.08,-0.04}$&$0.203^{+0.080,0.033}_{-0.080,-0.052}$&$0.13^{+0.39,+0.07}_{-0.13,-0.13}$&$0.00^{+0.01,+0.00}_{-0.00,-0.00}$&8.0\\
CL0024-iD1&$10.06^{+0.44,+0.28}_{-0.65,-0.21}$&$0.002^{+0.044,0.003}_{-0.001,-0.001}$&$1.61^{+0.53,+0.53}_{-1.45,-0.62}$&$0.02^{+0.02,+0.01}_{-0.02,-0.02}$&6.8\\
CL0024-zD1&$10.05^{+0.58,+0.38}_{-0.80,-0.30}$&$0.002^{+0.060,0.003}_{-0.002,-0.002}$&$1.78^{+0.86,+0.74}_{-1.53,-0.61}$&$0.03^{+0.02,+0.02}_{-0.02,-0.01}$&7.0\\

\enddata
\tablecomments{This table lists the median values for each parameters as well as the 95\% and 68\% credible regions, separated by a comma.\\
(a) Extinction \citep{calzetti2000}\\
(b) Metallicity (Solar metallicity is Z=0.02)\\
(c) Earliest formation redshift\\
}
\end{deluxetable}

\clearpage

\clearpage
\pagestyle{empty}
\begin{deluxetable}{ccccccc}
\tabletypesize{\footnotesize}
\setlength{\tabcolsep}{0.06in}
\tablewidth{0pt}\tablecaption{\PMCMC\ EXP results for object listes in Table \ref{phottable}\label{expres}}
\tablehead{\colhead{UID} & \colhead{Mass (${\rm M_\sun}$)}  & \colhead{Age (Gyr)}  & \colhead{\AV\tablenotemark{a}} & \colhead{Z\tablenotemark{c}}  & \colhead{$\tau$ (Gyr)} }
\startdata
631&$8.18^{+0.40,+0.18}_{-0.34,-0.22}$&$0.008^{+0.041,0.005}_{-0.008,-0.008}$&$0.34^{+0.40,+0.26}_{-0.32,-0.23}$&$0.03^{+0.02,+0.02}_{-0.02,-0.01}$&$6.51^{+6.44,+6.47}_{-5.88,-2.33}$\\
712&$8.17^{+0.57,+0.29}_{-0.53,-0.31}$&$0.002^{+0.055,0.003}_{-0.001,-0.001}$&$0.46^{+0.48,+0.26}_{-0.46,-0.32}$&$0.02^{+0.02,+0.01}_{-0.02,-0.02}$&$6.63^{+5.97,+5.95}_{-6.26,-2.75}$\\
4442&$7.27^{+0.87,+0.24}_{-0.55,-0.34}$&$0.001^{+0.195,0.002}_{-0.001,-0.001}$&$0.39^{+0.38,+0.22}_{-0.39,-0.27}$&$0.03^{+0.02,+0.02}_{-0.02,-0.01}$&$6.43^{+6.19,+2.52}_{-6.13,-6.26}$\\
5183&$7.66^{+0.66,+0.25}_{-0.54,-0.29}$&$0.002^{+0.137,0.005}_{-0.002,-0.002}$&$0.38^{+0.35,+0.19}_{-0.38,-0.30}$&$0.02^{+0.03,+0.01}_{-0.02,-0.02}$&$6.45^{+5.96,+2.53}_{-6.38,-6.29}$\\
5225&$8.83^{+0.43,+0.22}_{-0.39,-0.23}$&$0.005^{+0.065,0.008}_{-0.005,-0.005}$&$0.56^{+0.45,+0.28}_{-0.48,-0.33}$&$0.03^{+0.02,+0.02}_{-0.02,-0.01}$&$6.53^{+6.22,+5.11}_{-6.09,-3.65}$\\
6139&$8.78^{+0.26,+0.16}_{-0.37,-0.16}$&$0.001^{+0.003,0.001}_{-0.001,-0.001}$&$0.67^{+0.19,+0.12}_{-0.30,-0.10}$&$0.04^{+0.01,+0.01}_{-0.02,-0.01}$&$6.47^{+5.92,+2.38}_{-6.41,-6.44}$\\
9040&$8.68^{+0.95,+0.28}_{-0.60,-0.35}$&$0.004^{+0.413,0.015}_{-0.003,-0.003}$&$0.53^{+0.38,+0.33}_{-0.53,-0.31}$&$0.02^{+0.03,+0.01}_{-0.02,-0.02}$&$6.54^{+6.44,+5.67}_{-5.83,-3.07}$\\
9340&$7.49^{+1.29,+0.21}_{-0.26,-0.24}$&$0.002^{+0.690,0.016}_{-0.002,-0.002}$&$0.09^{+0.28,+0.06}_{-0.09,-0.09}$&$0.00^{+0.00,+0.00}_{-0.00,-0.00}$&$6.52^{+6.27,+4.58}_{-5.96,-4.07}$\\
9487&$8.21^{+0.62,+0.33}_{-0.61,-0.33}$&$0.040^{+0.223,0.042}_{-0.039,-0.039}$&$0.16^{+0.51,+0.12}_{-0.16,-0.16}$&$0.01^{+0.02,+0.00}_{-0.01,-0.01}$&$6.56^{+6.44,+4.03}_{-5.86,-4.71}$\\
L7-01&$9.79^{+0.67,+0.35}_{-0.66,-0.33}$&$0.046^{+0.571,0.109}_{-0.046,-0.046}$&$1.28^{+0.97,+0.59}_{-0.97,-0.49}$&$0.03^{+0.02,+0.02}_{-0.02,-0.01}$&$6.35^{+6.24,+5.28}_{-6.07,-3.45}$\\
L8-01&$8.70^{+0.84,+0.54}_{-0.91,-0.47}$&$0.002^{+0.212,0.007}_{-0.002,-0.002}$&$0.70^{+0.87,+0.29}_{-0.70,-0.66}$&$0.02^{+0.02,+0.01}_{-0.02,-0.02}$&$6.47^{+6.42,+2.92}_{-5.88,-5.90}$\\
A1689-zD1&$10.36^{+0.53,+0.31}_{-0.57,-0.29}$&$0.006^{+0.372,0.019}_{-0.006,-0.006}$&$1.41^{+0.76,+0.63}_{-1.11,-0.43}$&$0.02^{+0.03,+0.01}_{-0.02,-0.02}$&$6.50^{+6.49,+3.72}_{-5.81,-5.06}$\\
A1703-iD1&$10.46^{+0.57,+0.43}_{-0.75,-0.30}$&$0.154^{+0.689,0.147}_{-0.154,-0.154}$&$0.43^{+0.73,+0.19}_{-0.43,-0.43}$&$0.02^{+0.02,+0.01}_{-0.02,-0.02}$&$6.61^{+6.39,+6.39}_{-5.80,-2.29}$\\
A2218-iD1&$10.24^{+0.45,+0.24}_{-0.52,-0.25}$&$0.002^{+0.139,0.004}_{-0.002,-0.002}$&$1.42^{+0.57,+0.48}_{-1.04,-0.34}$&$0.03^{+0.02,+0.02}_{-0.03,-0.01}$&$6.47^{+6.09,+4.10}_{-6.25,-4.71}$\\
A383-iD1&$11.17^{+0.13,+0.08}_{-0.14,-0.06}$&$0.991^{+0.158,0.158}_{-0.326,-0.079}$&$0.24^{+0.47,+0.11}_{-0.24,-0.24}$&$0.01^{+0.03,+0.01}_{-0.01,-0.01}$&$0.44^{+2.76,+0.10}_{-0.28,-0.16}$\\
CL0024-iD1&$10.11^{+0.45,+0.26}_{-0.49,-0.24}$&$0.005^{+0.552,0.029}_{-0.005,-0.005}$&$1.37^{+0.73,+0.68}_{-1.23,-0.52}$&$0.02^{+0.02,+0.01}_{-0.02,-0.02}$&$6.61^{+6.29,+5.32}_{-6.00,-3.37}$\\
CL0024-zD1&$10.08^{+0.62,+0.35}_{-0.71,-0.30}$&$0.004^{+0.371,0.014}_{-0.004,-0.004}$&$1.73^{+0.91,+0.65}_{-1.20,-0.53}$&$0.03^{+0.02,+0.02}_{-0.02,-0.01}$&$6.44^{+6.06,+2.59}_{-6.23,-6.16}$\\
\enddata
\tablecomments{This table lists the median values for each parameters as well as the 95\% and 68\% credible regions, separated by a comma.\\
(a) Extinction \citep{calzetti2000}\\
(b) Metallicity (Solar metallicity is Z=0.02)\\
}
\end{deluxetable}

\clearpage

\clearpage
\pagestyle{empty}
\begin{deluxetable}{cccccccc}
\tabletypesize{\footnotesize}
\renewcommand{\arraystretch}{.6}
\setlength{\tabcolsep}{0.06in}
\tablewidth{0pt}
\tablewidth{0pt}\tablecaption{\PMCMC\ SSP2 results for object listes in Table \ref{phottable}\label{ssp2res}}
\tablehead{\colhead{UID} & \colhead{Mass (${\rm M_\sun}$)}  & \colhead{${\rm Age_{Old}}$ (Gyr)}  & \colhead{\AV\tablenotemark{a}} & \colhead{${\rm Z_{old}}$\tablenotemark{b}}  & \colhead{\% Old}  & \colhead{${\rm Age_{Young}}$ (Gyr)} & \colhead{${\rm Z_{young}}$\tablenotemark{b}}   }
\rotate
\startdata
631&$8.29^{+0.65,+0.27}_{-0.56,-0.42}$&$0.045^{+1.026,0.108}_{-0.045,-0.045}$&$0.47^{+0.25,+0.24}_{-0.46,-0.19}$&$0.03^{+0.02,+0.02}_{-0.02,-0.01}$&$44.69^{+43.55,+16.40}_{-44.61,-44.59}$&$0.001^{+0.004,0.001}_{-0.001,-0.001}$&$0.03^{+0.02,+0.02}_{-0.02,-0.01}$\\
712&$8.25^{+0.67,+0.32}_{-0.63,-0.36}$&$0.029^{+0.786,0.087}_{-0.029,-0.029}$&$0.46^{+0.44,+0.27}_{-0.46,-0.28}$&$0.02^{+0.02,+0.01}_{-0.02,-0.02}$&$42.68^{+47.34,+17.48}_{-42.68,-42.68}$&$0.001^{+0.004,0.001}_{-0.000,-0.000}$&$0.02^{+0.02,+0.01}_{-0.02,-0.02}$\\
4442&$7.24^{+1.02,+0.31}_{-0.65,-0.58}$&$0.043^{+0.628,0.079}_{-0.042,-0.042}$&$0.38^{+0.41,+0.16}_{-0.37,-0.32}$&$0.03^{+0.02,+0.02}_{-0.02,-0.01}$&$56.26^{+42.71,+42.34}_{-50.36,-19.56}$&$0.001^{+0.015,0.001}_{-0.001,-0.001}$&$0.03^{+0.02,+0.02}_{-0.02,-0.01}$\\
5183&$7.73^{+0.75,+0.30}_{-0.62,-0.34}$&$0.033^{+0.840,0.092}_{-0.033,-0.033}$&$0.38^{+0.33,+0.23}_{-0.38,-0.23}$&$0.02^{+0.02,+0.01}_{-0.02,-0.02}$&$46.87^{+45.54,+17.90}_{-46.87,-46.87}$&$0.001^{+0.006,0.001}_{-0.001,-0.001}$&$0.02^{+0.03,+0.01}_{-0.02,-0.02}$\\
5225&$8.82^{+0.55,+0.30}_{-0.56,-0.32}$&$0.046^{+0.737,0.098}_{-0.045,-0.045}$&$0.68^{+0.29,+0.34}_{-0.68,-0.25}$&$0.03^{+0.02,+0.02}_{-0.02,-0.01}$&$36.48^{+51.59,+19.68}_{-36.48,-36.48}$&$0.001^{+0.009,0.001}_{-0.001,-0.001}$&$0.03^{+0.02,+0.02}_{-0.03,-0.01}$\\
6139&$8.76^{+0.35,+0.20}_{-0.47,-0.19}$&$0.008^{+0.887,0.090}_{-0.008,-0.008}$&$0.62^{+0.25,+0.16}_{-0.37,-0.11}$&$0.03^{+0.02,+0.02}_{-0.03,-0.01}$&$31.47^{+58.81,+17.41}_{-31.47,-31.47}$&$0.001^{+0.002,0.001}_{-0.001,-0.001}$&$0.04^{+0.01,+0.01}_{-0.03,-0.01}$\\
9040&$8.71^{+0.76,+0.29}_{-0.65,-0.32}$&$0.034^{+0.811,0.092}_{-0.034,-0.034}$&$0.60^{+0.31,+0.26}_{-0.58,-0.19}$&$0.02^{+0.02,+0.01}_{-0.02,-0.02}$&$37.18^{+55.03,+21.66}_{-37.18,-37.18}$&$0.001^{+0.009,0.001}_{-0.000,-0.000}$&$0.02^{+0.03,+0.01}_{-0.02,-0.02}$\\
9340&$10.60^{+0.12,+0.07}_{-0.21,-0.07}$&$1.220^{+0.050,0.050}_{-0.141,-0.033}$&$0.03^{+0.07,+0.01}_{-0.03,-0.03}$&$0.05^{+0.00,+0.00}_{-0.00,-0.00}$&$99.95^{+0.00,+0.01}_{-0.04,-0.01}$&$0.001^{+0.002,0.000}_{-0.000,-0.000}$&$0.00^{+0.00,+0.00}_{-0.00,-0.00}$\\
9487&$8.40^{+0.58,+0.32}_{-0.79,-0.35}$&$0.068^{+0.985,0.114}_{-0.068,-0.068}$&$0.42^{+0.31,+0.24}_{-0.42,-0.23}$&$0.02^{+0.02,+0.01}_{-0.02,-0.02}$&$55.88^{+40.01,+40.01}_{-51.33,-22.10}$&$0.000^{+0.002,0.000}_{-0.000,-0.000}$&$0.01^{+0.03,+0.01}_{-0.01,-0.01}$\\
L7-01&$9.83^{+0.60,+0.34}_{-0.77,-0.32}$&$0.080^{+0.509,0.090}_{-0.080,-0.080}$&$1.39^{+0.92,+0.65}_{-1.33,-0.71}$&$0.02^{+0.02,+0.02}_{-0.02,-0.01}$&$53.06^{+40.35,+40.81}_{-53.03,-23.79}$&$0.003^{+0.068,0.009}_{-0.003,-0.003}$&$0.02^{+0.02,+0.01}_{-0.02,-0.02}$\\
L8-01&$8.74^{+0.84,+0.55}_{-0.94,-0.49}$&$0.026^{+0.423,0.060}_{-0.026,-0.026}$&$0.70^{+0.80,+0.37}_{-0.70,-0.62}$&$0.03^{+0.02,+0.01}_{-0.03,-0.03}$&$45.37^{+47.53,+19.44}_{-44.96,-44.03}$&$0.001^{+0.012,0.001}_{-0.000,-0.000}$&$0.02^{+0.02,+0.01}_{-0.02,-0.02}$\\
A1689-zD1&$10.35^{+0.63,+0.33}_{-0.69,-0.29}$&$0.043^{+0.654,0.090}_{-0.043,-0.043}$&$1.40^{+0.71,+0.57}_{-1.12,-0.43}$&$0.02^{+0.02,+0.01}_{-0.02,-0.02}$&$47.71^{+45.93,+17.46}_{-47.66,-47.68}$&$0.001^{+0.022,0.002}_{-0.001,-0.001}$&$0.02^{+0.02,+0.01}_{-0.02,-0.02}$\\
A1703-iD1&$10.37^{+0.60,+0.36}_{-0.82,-0.31}$&$0.109^{+0.781,0.129}_{-0.109,-0.109}$&$0.31^{+0.90,+0.24}_{-0.31,-0.31}$&$0.02^{+0.02,+0.01}_{-0.02,-0.02}$&$59.62^{+40.05,+40.23}_{-53.59,-20.13}$&$0.010^{+0.035,0.008}_{-0.010,-0.010}$&$0.02^{+0.02,+0.01}_{-0.02,-0.02}$\\
A2218-iD1&$10.24^{+0.47,+0.31}_{-0.71,-0.25}$&$0.046^{+0.798,0.106}_{-0.046,-0.046}$&$1.43^{+0.57,+0.49}_{-1.14,-0.30}$&$0.02^{+0.02,+0.01}_{-0.02,-0.02}$&$1.00^{+67.18,+5.33}_{-1.00,-1.00}$&$0.001^{+0.014,0.001}_{-0.001,-0.001}$&$0.03^{+0.02,+0.02}_{-0.02,-0.01}$\\
A383-iD1&$11.32^{+0.30,+0.16}_{-0.31,-0.16}$&$0.458^{+0.498,0.157}_{-0.307,-0.225}$&$0.56^{+0.37,+0.24}_{-0.55,-0.21}$&$0.03^{+0.02,+0.02}_{-0.02,-0.01}$&$99.15^{+0.57,+0.37}_{-2.62,-0.30}$&$0.000^{+0.011,0.000}_{-0.000,-0.000}$&$0.01^{+0.03,+0.01}_{-0.01,-0.01}$\\
CL0024-iD1&$10.14^{+0.55,+0.28}_{-0.64,-0.27}$&$0.048^{+0.829,0.112}_{-0.048,-0.048}$&$1.53^{+0.57,+0.52}_{-1.34,-0.43}$&$0.02^{+0.02,+0.01}_{-0.02,-0.02}$&$44.99^{+48.91,+17.91}_{-44.99,-44.99}$&$0.001^{+0.023,0.001}_{-0.001,-0.001}$&$0.02^{+0.02,+0.01}_{-0.02,-0.02}$\\
CL0024-zD1&$10.13^{+0.68,+0.37}_{-0.82,-0.33}$&$0.040^{+0.801,0.103}_{-0.040,-0.040}$&$1.80^{+0.81,+0.61}_{-1.37,-0.47}$&$0.03^{+0.02,+0.02}_{-0.02,-0.01}$&$44.44^{+48.11,+17.67}_{-44.44,-44.44}$&$0.001^{+0.025,0.001}_{-0.001,-0.001}$&$0.03^{+0.02,+0.02}_{-0.03,-0.01}$\\
\enddata
\tablecomments{This table lists the median values for each parameters as well as the 95\% and 68\% credible regions, separated by a comma.\\
(a) Extinction \citep{calzetti2000}\\
(b) Metallicity (Solar metallicity is Z=0.02)\\
}
\end{deluxetable}

\clearpage

\clearpage
\pagestyle{empty}
\begin{deluxetable}{ccccccc}
\tabletypesize{\footnotesize}
\setlength{\tabcolsep}{0.06in}
\tablewidth{0pt}
\tablewidth{0pt}\tablecaption{\PMCMC\ Maximum M/L SSP2 results for object listes in Table \ref{phottable}\label{mupres}}
\tablehead{\colhead{UID} & \colhead{Mass (${\rm M_\sun}$)}   & \colhead{\AV\tablenotemark{a}} & \colhead{${\rm Z_{old}}$\tablenotemark{b}}  & \colhead{\% Old}  & \colhead{${\rm Age_{Young}}$ (Gyr)} & \colhead{${\rm Z_{young}}$\tablenotemark{b}}   }
\rotate
\startdata
631&$8.50^{+0.82,+0.40}_{-0.69,-0.51}$&$0.51^{+0.24,+0.26}_{-0.51,-0.24}$&$0.02^{+0.02,+0.01}_{-0.02,-0.02}$&$54.60^{+41.39,+41.38}_{-49.73,-19.46}$&$0.001^{+0.004,0.002}_{-0.001,-0.001}$&$0.03^{+0.02,+0.02}_{-0.02,-0.02}$\\
712&$8.50^{+0.79,+0.38}_{-0.75,-0.42}$&$0.51^{+0.45,+0.27}_{-0.51,-0.29}$&$0.02^{+0.03,+0.01}_{-0.02,-0.02}$&$45.14^{+44.24,+16.84}_{-45.13,-45.01}$&$0.001^{+0.008,0.001}_{-0.001,-0.001}$&$0.01^{+0.03,+0.01}_{-0.01,-0.01}$\\
4442&$7.57^{+1.05,+0.36}_{-0.75,-0.48}$&$0.44^{+0.36,+0.21}_{-0.44,-0.23}$&$0.03^{+0.02,+0.01}_{-0.02,-0.02}$&$48.23^{+45.70,+18.04}_{-48.23,-48.23}$&$0.001^{+0.005,0.001}_{-0.000,-0.000}$&$0.03^{+0.02,+0.02}_{-0.02,-0.01}$\\
5183&$7.97^{+1.04,+0.38}_{-0.78,-0.45}$&$0.45^{+0.31,+0.23}_{-0.45,-0.20}$&$0.02^{+0.02,+0.01}_{-0.02,-0.02}$&$48.16^{+46.35,+18.17}_{-48.16,-48.16}$&$0.001^{+0.009,0.001}_{-0.001,-0.001}$&$0.02^{+0.03,+0.01}_{-0.02,-0.02}$\\
5225&$9.07^{+0.77,+0.32}_{-0.67,-0.50}$&$0.67^{+0.31,+0.35}_{-0.67,-0.29}$&$0.03^{+0.02,+0.02}_{-0.02,-0.01}$&$47.41^{+42.97,+28.09}_{-47.20,-36.20}$&$0.001^{+0.008,0.002}_{-0.001,-0.001}$&$0.03^{+0.02,+0.02}_{-0.03,-0.01}$\\
6139&$8.79^{+0.40,+0.20}_{-0.46,-0.22}$&$0.64^{+0.23,+0.14}_{-0.32,-0.12}$&$0.03^{+0.02,+0.02}_{-0.02,-0.01}$&$14.27^{+51.70,+15.94}_{-14.27,-14.27}$&$0.001^{+0.002,0.001}_{-0.001,-0.001}$&$0.04^{+0.01,+0.01}_{-0.02,-0.01}$\\
9040&$8.76^{+0.90,+0.31}_{-0.75,-0.38}$&$0.64^{+0.29,+0.27}_{-0.64,-0.17}$&$0.03^{+0.02,+0.02}_{-0.02,-0.01}$&$29.35^{+61.96,+24.20}_{-29.35,-29.35}$&$0.001^{+0.012,0.001}_{-0.001,-0.001}$&$0.02^{+0.03,+0.01}_{-0.02,-0.02}$\\
9340&$10.64^{+0.14,+0.08}_{-0.22,-0.07}$&$0.02^{+0.08,+0.02}_{-0.02,-0.02}$&$0.05^{+0.00,+0.00}_{-0.00,-0.00}$&$99.95^{+0.02,+0.01}_{-0.02,-0.00}$&$0.001^{+0.002,0.000}_{-0.000,-0.000}$&$0.00^{+0.00,+0.00}_{-0.00,-0.00}$\\
9487&$8.44^{+0.95,+0.39}_{-0.75,-0.48}$&$0.54^{+0.27,+0.27}_{-0.52,-0.21}$&$0.02^{+0.02,+0.01}_{-0.02,-0.02}$&$52.75^{+44.43,+44.06}_{-48.37,-20.32}$&$0.001^{+0.019,0.003}_{-0.001,-0.001}$&$0.01^{+0.03,+0.00}_{-0.01,-0.01}$\\
L7-01&$9.99^{+0.60,+0.36}_{-0.86,-0.31}$&$1.06^{+1.18,+0.39}_{-1.05,-1.00}$&$0.03^{+0.02,+0.02}_{-0.03,-0.01}$&$43.27^{+47.42,+16.55}_{-43.25,-43.25}$&$0.012^{+0.076,0.018}_{-0.012,-0.012}$&$0.02^{+0.03,+0.01}_{-0.02,-0.02}$\\
L8-01&$8.66^{+0.75,+0.53}_{-0.93,-0.47}$&$0.76^{+0.83,+0.41}_{-0.75,-0.63}$&$0.03^{+0.02,+0.01}_{-0.02,-0.02}$&$0.11^{+0.82,+0.11}_{-0.11,-0.11}$&$0.001^{+0.021,0.001}_{-0.001,-0.001}$&$0.02^{+0.02,+0.01}_{-0.02,-0.02}$\\
A1689-zD1&$10.54^{+0.82,+0.42}_{-0.79,-0.37}$&$1.32^{+0.78,+0.65}_{-1.14,-0.52}$&$0.03^{+0.02,+0.02}_{-0.02,-0.01}$&$48.19^{+44.97,+29.13}_{-48.13,-36.78}$&$0.002^{+0.033,0.004}_{-0.002,-0.002}$&$0.02^{+0.02,+0.01}_{-0.02,-0.02}$\\
A1703-iD1&$10.52^{+0.75,+0.43}_{-0.86,-0.37}$&$0.41^{+0.86,+0.18}_{-0.41,-0.41}$&$0.02^{+0.02,+0.01}_{-0.02,-0.02}$&$51.71^{+46.05,+44.77}_{-47.56,-20.92}$&$0.012^{+0.033,0.007}_{-0.012,-0.012}$&$0.02^{+0.03,+0.01}_{-0.02,-0.02}$\\
A2218-iD1&$10.46^{+0.69,+0.36}_{-0.71,-0.32}$&$1.39^{+0.55,+0.48}_{-1.18,-0.33}$&$0.02^{+0.02,+0.01}_{-0.02,-0.02}$&$42.59^{+46.42,+17.00}_{-42.59,-42.59}$&$0.001^{+0.016,0.001}_{-0.001,-0.001}$&$0.03^{+0.02,+0.02}_{-0.02,-0.01}$\\
A383-iD1&$11.47^{+0.25,+0.11}_{-0.23,-0.13}$&$0.48^{+0.46,+0.29}_{-0.48,-0.32}$&$0.02^{+0.03,+0.01}_{-0.02,-0.02}$&$99.34^{+0.47,+0.20}_{-2.28,-0.24}$&$0.001^{+0.023,0.006}_{-0.001,-0.001}$&$0.01^{+0.03,+0.01}_{-0.01,-0.01}$\\
CL0024-iD1&$10.29^{+0.73,+0.37}_{-0.71,-0.31}$&$1.55^{+0.54,+0.54}_{-1.39,-0.53}$&$0.02^{+0.02,+0.01}_{-0.02,-0.02}$&$44.00^{+47.07,+17.71}_{-44.00,-44.00}$&$0.002^{+0.032,0.003}_{-0.001,-0.001}$&$0.02^{+0.02,+0.01}_{-0.02,-0.02}$\\
CL0024-zD1&$10.27^{+0.77,+0.40}_{-0.87,-0.38}$&$1.77^{+0.82,+0.70}_{-1.52,-0.54}$&$0.02^{+0.02,+0.01}_{-0.02,-0.02}$&$34.14^{+54.14,+19.25}_{-34.14,-34.14}$&$0.002^{+0.049,0.002}_{-0.002,-0.002}$&$0.03^{+0.02,+0.02}_{-0.03,-0.01}$\\
\enddata
\tablecomments{This table lists the median values for each parameters as well as the 95\% and 68\% credible regions, separated by a comma.\\
(a) Extinction \citep{calzetti2000}\\
(b) Metallicity (Solar metallicity is Z=0.02)\\
}
\end{deluxetable}

\clearpage

\clearpage
\pagestyle{empty}
\begin{deluxetable}{ccccccc}
\tabletypesize{\footnotesize}
\setlength{\tabcolsep}{0.06in}
\tablewidth{0pt}\tablecaption{\PMCMC\ SSP+Nebular results for object listes in Table \ref{phottable}\label{nebres}}
\tablehead{\colhead{UID} & \colhead{Mass (${\rm M_\sun}$)}  & \colhead{Age (Gyr)}  & \colhead{\AV\tablenotemark{a}} & \colhead{Z\tablenotemark{b}}  & \colhead{$f_{esc}$\tablenotemark{c}} & \colhead{${\rm z_f}$\tablenotemark{d}}}
\startdata
631&$8.48^{+0.24,+0.15}_{-0.34,-0.10}$&$0.001^{+0.012,0.001}_{-0.001,-0.001}$&$0.66^{+0.20,+0.14}_{-0.41,-0.09}$&$0.00^{+0.01,+0.00}_{-0.00,-0.00}$&$0.95^{+0.05,+0.05}_{-0.63,-0.03}$&4.0\\
712&$8.00^{+0.46,+0.27}_{-0.48,-0.25}$&$0.001^{+0.002,0.000}_{-0.000,-0.000}$&$0.45^{+0.37,+0.24}_{-0.42,-0.21}$&$0.02^{+0.02,+0.01}_{-0.02,-0.02}$&$0.41^{+0.31,+0.23}_{-0.40,-0.20}$&5.2\\
4442&$6.90^{+0.29,+0.09}_{-0.18,-0.15}$&$0.000^{+0.001,0.000}_{-0.000,-0.000}$&$0.14^{+0.21,+0.05}_{-0.14,-0.14}$&$0.04^{+0.01,+0.01}_{-0.04,-0.01}$&$0.06^{+0.19,+0.04}_{-0.06,-0.06}$&5.8\\
5183&$7.13^{+0.08,+0.03}_{-0.06,-0.04}$&$0.000^{+0.000,0.000}_{-0.000,-0.000}$&$0.05^{+0.09,+0.02}_{-0.05,-0.05}$&$0.05^{+0.00,+0.00}_{-0.01,-0.00}$&$0.04^{+0.16,+0.04}_{-0.04,-0.04}$&4.8\\
5225&$8.57^{+0.40,+0.20}_{-0.45,-0.22}$&$0.001^{+0.005,0.001}_{-0.001,-0.001}$&$0.55^{+0.26,+0.17}_{-0.51,-0.15}$&$0.03^{+0.02,+0.02}_{-0.02,-0.01}$&$0.58^{+0.37,+0.32}_{-0.54,-0.21}$&5.4\\
6139&$8.15^{+0.19,+0.07}_{-0.15,-0.09}$&$0.003^{+0.001,0.000}_{-0.001,-0.001}$&$0.14^{+0.20,+0.07}_{-0.14,-0.14}$&$0.04^{+0.01,+0.01}_{-0.02,-0.00}$&$0.74^{+0.26,+0.26}_{-0.50,-0.12}$&4.9\\
9040&$8.33^{+0.19,+0.10}_{-0.18,-0.11}$&$0.000^{+0.001,0.000}_{-0.000,-0.000}$&$0.34^{+0.16,+0.08}_{-0.14,-0.08}$&$0.00^{+0.03,+0.00}_{-0.00,-0.00}$&$0.09^{+0.20,+0.05}_{-0.09,-0.09}$&4.9\\
9340&$7.29^{+0.04,+0.01}_{-0.02,-0.01}$&$0.000^{+0.001,0.000}_{-0.000,-0.000}$&$0.01^{+0.02,+0.00}_{-0.01,-0.01}$&$0.00^{+0.00,+0.00}_{-0.00,-0.00}$&$0.00^{+0.06,+0.01}_{-0.00,-0.00}$&4.7\\
9487&$7.65^{+0.54,+0.21}_{-0.41,-0.26}$&$0.001^{+0.010,0.001}_{-0.001,-0.001}$&$0.26^{+0.31,+0.14}_{-0.26,-0.16}$&$0.02^{+0.02,+0.01}_{-0.02,-0.02}$&$0.47^{+0.43,+0.30}_{-0.46,-0.29}$&4.1\\
L7-01&$9.34^{+0.76,+0.38}_{-0.58,-0.36}$&$0.001^{+0.087,0.003}_{-0.001,-0.001}$&$1.41^{+0.60,+0.43}_{-1.18,-0.35}$&$0.02^{+0.03,+0.01}_{-0.02,-0.02}$&$0.00^{+0.70,+0.02}_{-0.00,-0.00}$&7.7\\
L8-01&$8.55^{+0.84,+0.62}_{-0.90,-0.38}$&$0.003^{+0.037,0.005}_{-0.002,-0.002}$&$0.52^{+0.81,+0.24}_{-0.52,-0.51}$&$0.03^{+0.02,+0.02}_{-0.02,-0.01}$&$0.61^{+0.38,+0.39}_{-0.54,-0.20}$&8.4\\
A1689-zD1&$10.01^{+0.53,+0.35}_{-0.61,-0.30}$&$0.003^{+0.033,0.005}_{-0.003,-0.003}$&$1.07^{+0.74,+0.49}_{-0.88,-0.38}$&$0.03^{+0.02,+0.02}_{-0.02,-0.01}$&$0.54^{+0.46,+0.46}_{-0.48,-0.19}$&7.9\\
A1703-iD1&$9.86^{+0.76,+0.48}_{-0.66,-0.43}$&$0.004^{+0.042,0.008}_{-0.004,-0.004}$&$0.53^{+0.54,+0.28}_{-0.53,-0.35}$&$0.02^{+0.03,+0.01}_{-0.02,-0.02}$&$0.49^{+0.44,+0.28}_{-0.48,-0.35}$&6.2\\
A2218-iD1&$9.88^{+0.60,+0.42}_{-0.61,-0.33}$&$0.005^{+0.030,0.005}_{-0.005,-0.005}$&$0.67^{+0.74,+0.30}_{-0.66,-0.43}$&$0.03^{+0.02,+0.02}_{-0.03,-0.01}$&$0.56^{+0.43,+0.43}_{-0.52,-0.21}$&6.9\\
A383-iD1&$9.75^{+0.17,+0.08}_{-0.15,-0.08}$&$0.000^{+0.001,0.000}_{-0.000,-0.000}$&$1.04^{+0.16,+0.07}_{-0.14,-0.08}$&$0.00^{+0.01,+0.00}_{-0.00,-0.00}$&$0.08^{+0.23,+0.06}_{-0.08,-0.08}$&6.0\\
CL0024-iD1&$9.79^{+0.50,+0.40}_{-0.64,-0.30}$&$0.005^{+0.044,0.007}_{-0.004,-0.004}$&$0.84^{+0.90,+0.48}_{-0.83,-0.48}$&$0.02^{+0.02,+0.01}_{-0.02,-0.02}$&$0.60^{+0.40,+0.40}_{-0.54,-0.20}$&6.8\\
CL0024-zD1&$9.79^{+0.69,+0.45}_{-0.80,-0.39}$&$0.004^{+0.069,0.007}_{-0.004,-0.004}$&$1.25^{+0.87,+0.63}_{-1.19,-0.49}$&$0.03^{+0.02,+0.02}_{-0.02,-0.01}$&$0.54^{+0.46,+0.46}_{-0.49,-0.20}$&7.1\\
\enddata
\tablecomments{This table lists the median values for each parameters as well as the 95\% and 68\% credible regions, separated by a comma..\\
(a) Extinction \citep{calzetti2000}\\
(b) Metallicity (Solar metallicity is Z=0.02)\\
(c) Ionizing radiation escape fraction\\
(d) Earliest formation redshift\\
}
\end{deluxetable}

\clearpage

\clearpage
\pagestyle{empty}
\begin{deluxetable}{cccccccc}
\tablecaption{\PMCMC\ SSP results using MA05 and BC03 models for a subsample of sources from Table \ref{phottable} and allowing only stellar mass, stellar ages and extinction to vary.\label{GRAPESSSPMA05}}
\tablehead{
\colhead{} & \multicolumn{3}{c}{MA05} &  \colhead{} & \multicolumn{3}{c}{BC03}\\
\cline{2-4} \cline{6-8} \\ 
\colhead{UID} & \colhead{Age (Gyr)} & \colhead{\AV\tablenotemark{a}} & \colhead{Mass (${\rm M_\sun}$)}  &  \colhead{} & \colhead{Age (Gyr)} & \colhead{\AV\tablenotemark{a}} & \colhead{Mass (${\rm M_\sun}$)} }
\startdata
4442 & $0.022_{-0.02}^{+0.05}$  & $0.22_{-0.22}^{+0.56}$  & $7.74_{-0.79}^{+0.61}$ & & $0.045_{-0.05}^{0.11}$  & $0.28_{-0.28}^{0.54}$  & $8.20_{-0.66}^{0.57}$\\
5183 & $0.013_{-0.01}^{+0.03}$  & $0.22_{-0.22}^{+0.56}$  & $7.86_{-0.57}^{+0.52}$ & &  $0.023_{-0.02}^{0.06}$  & $0.35_{-0.35}^{0.45}$  & $8.32_{-0.52}^{0.40}$  \\
5225 & $0.007_{-0.01}^{+0.01}$  & $0.58_{-0.32}^{+0.51}$  & $8.83_{-0.38}^{+0.35}$ & &  $0.016_{-0.02}^{0.01}$  & $0.73_{-0.38}^{0.36}$  & $9.44_{-0.35}^{0.22}$\\
6139 & $0.002_{-0.00}^{+0.00}$  & $0.81_{-0.27}^{+0.23}$  & $8.95_{-0.34}^{+0.23}$ & &  $0.008_{-0.01}^{0.01}$  & $0.82_{-0.26}^{0.27}$  & $9.32_{-0.23}^{0.23}$\\
9040 & $0.002_{-0.00}^{+0.01}$  & $0.65_{-0.54}^{+0.26}$  & $8.49_{-0.38}^{+0.34}$ & &  $0.009_{-0.01}^{0.01}$  & $0.64_{-0.29}^{0.30}$  & $8.91_{-0.30}^{0.28}$\\
9340 & $0.017_{-0.02}^{+0.03}$  & $0.22_{-0.22}^{+0.61}$  & $8.08_{-0.58}^{+0.47}$ & & $0.023_{-0.02}^{0.07}$  & $0.40_{-0.40}^{0.43}$  & $8.48_{-0.56}^{0.39}$\\
9487 & $0.016_{-0.02}^{+0.02}$  & $0.17_{-0.17}^{+0.50}$  & $8.12_{-0.53}^{+0.36}$ & &  $0.036_{-0.04}^{0.03}$  & $0.20_{-0.20}^{0.46}$  & $8.66_{-0.34}^{0.21}$\\
\enddata
\tablecomments{This table lists the median values for each parameters as well as the 95\% credible regions.\\
(a) Extinction \citep{calzetti2000}\\
}
\end{deluxetable}

\clearpage

%===============================================================================
%%FIGURES

%\begin{figure}
%\includegraphics[width=5.0in]{newfig4.pdf}
%\caption{These plots show the approximate maximum likelihood function, as a function of extinction Av (marginalizing over all other model parameters) when the metallicity is added to the list of input parameters that are allowed to vary. The input data are the same as what was used in Figure \ref{1percent}.
%\label{likelihoodZ}}
%\end{figure}

\clearpage

\clearpage

\clearpage

\end{document}